\documentclass[12pt]{article}
\usepackage[margin=1in]{geometry}
\usepackage[switch]{lineno}
\usepackage[authoryear,round]{natbib}
\usepackage{graphicx}
\usepackage{booktabs}
\usepackage{longtable}
\usepackage{caption}
\usepackage[hidelinks]{hyperref}
\usepackage{cancel}

\usepackage{xr}
\makeatletter
\newcommand*{\addFileDependency}[1]{
  \typeout{(#1)}
  \@addtofilelist{#1}
  \IfFileExists{#1}{}{\typeout{No file #1.}}
}
\makeatother

\usepackage{natbib}
\usepackage{graphicx,verbatim,array,multicol, xcolor, lscape, mathrsfs}
\usepackage{psfrag, amsmath, amsfonts, epsfig, fancybox, setspace,soul,hyperref, float,booktabs}
\usepackage{amsthm}
\usepackage{caption}
\usepackage{subcaption}
\usepackage{multirow}
\newtheorem{theorem}{Theorem}
\newcommand{\keywords}[1]{\par\noindent\textbf{Keywords: }#1}

\def\supone{^{(1)}}
\def\supzero{^{(0)}}
\def\supg{^{(g)}}

\def\bX{\boldsymbol{X}}
\def\bZ{\boldsymbol{Z}}

\def\bY{\boldsymbol{Y}}
\def\bS{\boldsymbol{S}}
\def\bs{\boldsymbol{s}}
\def\btheta{\boldsymbol{\theta}}
\def\bepsilon{\boldsymbol{\epsilon}}
\def\bomega{\boldsymbol{\omega}}
\def\PTE{\text{PTE}}
\def\CPTE{\text{CPTE}}
\def\LPTE{\text{LPTE}}

\newcommand{\figureend}{1}
\newcommand{\blind}{0}

\newcommand{\makeblind}[1]{\if\blind0
#1
\else
[Hidden for anonymity]
\fi}

\title{\vspace{-1.5em}A Causal Framework for Evaluating Jointly Longitudinal Outcomes and Surrogate Markers: A State-Space Approach}
\author{
\normalsize
\makeblind{Silvaneo V. dos Santos Jr.$^{a}$ and Layla Parast$^{a}$\\[0.5em]
\normalsize $^{a}$Department of Statistics and Data Sciences, University of Texas at Austin}
}
\date{}

\begin{document}
\maketitle
\thispagestyle{empty}

\clearpage

\clearpage
\begin{abstract}
Surrogate markers offer the potential to reduce the burden of data collection by replacing costly or invasive primary outcomes with more accessible measurements, provided that they can faithfully indicate the effectiveness of a treatment. However, appropriate evaluation of a surrogate is particularly complex in longitudinal studies, where both outcomes and surrogates can evolve dynamically over time and interest lies not only in the treatment effect at one time, but rather treatment effects that may vary along the entire trajectory.  In this paper, we develop a statistical framework for surrogate evaluation when both the surrogate and primary outcome are measured over time. Specifically, within the potential outcomes framework, we propose a formal causal definition of the proportion of the treatment effect on the longitudinal primary outcome that is explained by the treatment effect on the longitudinal surrogate. For estimation, we leverage state-space models, together with the Kalman filter and smoother, enabling efficient estimation of treatment effects under realistic conditions of temporal evolution and patient-level variability. We introduce a nonparametric bootstrap strategy for state-space models, a temporal homogeneity test, and demonstrate the finite-sample performance of our proposed methods via a simulation study and application to a diabetes clinical trial.
\end{abstract}



\section{Introduction}

In clinical and biomedical research, the identification of valid surrogate markers is a central challenge. Surrogates offer the potential to reduce the burden of data collection by replacing costly or invasive primary outcomes with more accessible measurements, provided that they can faithfully indicate the effectiveness of a treatment. Several frameworks and methods to evaluate surrogate markers have been proposed with applications not only in biomedical research but also in social sciences and education \citep{elliott2023surrogate,blette2023low,Chetty2014Education,Athey2019Index}.

Surrogate evaluation is particularly complex in longitudinal studies, where both outcomes and surrogates can evolve dynamically over time and interest lies not only in the treatment effect at some single endpoint, but rather treatment effects that may vary along the entire trajectory. For example, in our motivating application using data from the Diabetes Control and Complications Trial (DCCT) study, we aim to evaluate hemoglobin A1c (HbA1c) as a surrogate for albumin excretion rate (AER), both of which are measured repeatedly over time. Previous work has investigated the validation of surrogates for longitudinal outcomes \citep{Roberts2022Longitudinal} and, separately, the validation of longitudinal surrogates, for both survival and non-survival outcomes \citep{Agniel2024Survival,Agniel2020Longitudinal}.
However, settings involving jointly longitudinal surrogates and outcomes have received limited attention in the literature. As a result, our goal of evaluating longitudinal HbA1c as a surrogate for longitudinal AER simply cannot be achieved using existing approaches.


\subsection{Our Contributions}
In this paper, we develop a statistical framework for evaluating a surrogate marker in longitudinal settings where both the surrogate and primary outcome are measured over time. Specifically, within the potential outcomes framework, we propose a formal causal definition of the proportion of the treatment effect on the longitudinal primary outcome that is explained by the treatment effect on the longitudinal surrogate (PTE). We consider treatment effect quantities that are both changes from baseline and  differences across entire trajectories, and formalize the conditions for identification of the causal quantities.  With respect to estimation, we consider the sequential nature of our longitudinal study setting and leverage state-space models \citep{West-DLM}, together with the Kalman filter and smoother \citep{Kalman_filter_origins}. State-space models are well established in the statistical literature; however, to the best of our knowledge, their application to surrogate evaluation has not previously been explored.  Our proposed framework enables efficient estimation of treatment effects under realistic conditions of temporal evolution, measurement error, and patient-level variability. For uncertainty quantification, we introduce a nonparametric bootstrap strategy specific to this setting that significantly reduces the computational burden of resampling procedures for replicated time-series data, thereby making large-scale uncertainty quantification feasible.

This article is structured as follows. In Section \ref{sec:causal}, we introduce notation and formally describe our proposed causal framework. In Section \ref{sec:estimating}, we present our proposed state-space approach to construct an estimator for the PTE, introduce a hypothesis test for temporal homogeneity of the PTE, and detail our proposed bootstrap approach for uncertainty quantification. Section \ref{sec:simul} reports simulation studies that assess the robustness of the framework across several scenarios and compare our method with natural alternatives. Section \ref{sec:app} illustrates the approach with an application to the DCCT clinical trial. Finally, in Section \ref{sec:conclusion}, we summarize our contributions and briefly discuss future work.

\vspace*{-5mm}
\section{Proposed Causal Framework}\label{sec:causal}

In this section, we introduce notation and formally present a causal framework for surrogate evaluation when both the outcome and the surrogate are longitudinal. We define the PTE in this context and highlight the specific challenges and considerations that the longitudinal setting introduces for surrogate evaluation. 
\subsection{Notation}\label{subsec:setting}

Let $Y_t$ denote the primary outcome measured longitudinally at discrete time points $t = 0, 1, \ldots, T$, with $\bY = (Y_0,\dots,Y_T)^{\top}$, $Y_{t} \in \mathbb{R}, \forall t$, representing the trajectory of the outcome, and $T$ denotes a fixed final time. Similarly, let $\bS$ denote a surrogate marker also measured at the same time points, with $\bS = (S_0,\dots,S_T)^{\top}$, $S_{t}  \in \mathbb{R}, \forall t$. Without loss of generality we assume higher values of $S$ and $T$ indicate ``better" outcomes. We denote the trajectory of the outcome up to time $t$ as $\bY_t = (Y_0,\dots,Y_t)^{\top}$ and, similarly, the trajectory of the surrogate up to time $t$ as $\bS_t = (S_0,\dots,S_t)^{\top}$. In this setting, $\bY$ is a measurement that is either more expensive or more invasive to the patient to measure, compared to $\bS$.  For example, in our diabetes example, $\bY$ is a measurement that requires an additional biological test whereas $\bS$ does not. Let $G \in \{0, 1\}$ be a randomized binary treatment indicator, where $G = 1$ denotes treatment and $G = 0$ denotes control.  We assume the treatment is randomly assigned at baseline ($t = 0$) and treatment assignment does not change during the study.  We adopt potential outcomes notation and denote $\bY^{(g)}$ as the potential outcome trajectory under treatment $g \in \{0,1\}$. That is, $\bY\supone$ and $\bY\supzero$ represent the longitudinal outcome trajectories under treatment and control, respectively. We first require a number of common causal conditions that are assumed to hold throughout:

\begin{enumerate} 
\item[] \textbf{(D1)} Consistency: $\bY\supg = \bY$ and $\bS\supg = \bS$ when $G = g$;
\item[] \textbf{(D2)} Positivity/Overlap:
$0<P(G = 1)<1$ for all subjects;
\item[] \textbf{(D3)} Unconfoundedness: $\bY\supg, \bS\supg \perp G$;
\end{enumerate} \vspace{-2mm}

\noindent Condition \textbf{(D1)} states that the observed outcome and surrogate under treatment $g$ are equal to their potential outcomes when treatment $G=g$ is actually received. Condition \textbf{(D2)} ensures overlap in treatment assignment while Condition \textbf{(D3)} requires that treatment assignment be independent of both the potential outcomes and the potential surrogate values. Conditions \textbf{(D2)} and \textbf{(D3)} are satisfied in our randomized trial setting.

\subsection{Definitions and Proposed PTE \label{subsec:PTE}}
In a non-longitudinal setting with a single surrogate marker and single primary outcome, \citet{wang2002measure} proposed a definition for the proportion of the treatment effect on the primary outcome that is explained by the treatment effect on the surrogate marker, commonly shortened to the proportion of treatment effect explained or PTE. Their approach involved defining a contrast between the overall treatment effect on $Y,$ and the \textit{residual treatment effect} on $Y$, which can be interpreted as the leftover treatment effect on $Y$ after accounting for the treatment effect on $S$. We build from this definition, extending it to our more complex setting with a longitudinal surrogate and outcome. To that end, we first define the \textit{global treatment effect} as the treatment effect on the entire outcome trajectory:
$$
\Delta_{\text{area}} = \sum_{t=0}^{T}  \mathbb{E}[Y\supone_t- Y\supzero_t].
$$

\noindent In addition, we define the \textit{local treatment effect} as $\Delta(t)=\mathbb{E}[Y\supone_t - Y\supzero_t]$. Next, we define the \textit{global residual treatment effect} as

$$
\Delta_{R,\text{area}} = \sum_{t=0}^{T}\int  \mathbb{E}[Y\supone_t- Y\supzero_t| \bS\supone_t = \bS\supzero_t=\boldsymbol{s}_t]dF_{\bS\supzero_t}(\boldsymbol{s}_t),
$$
where $F_{\bS\supzero}(\boldsymbol{s}_t)$ is the marginal cumulative distribution of the trajectory $\bS\supzero$. Note here that $\bS\supone_t$ and $\bS\supzero_t$ refer specifically to the trajectory of the surrogate up to time $t$, that is to say:

$$
\bS\supone_t = \bS\supzero_t=\boldsymbol{s}_t \iff S_h\supzero = S_h\supone =s_h, \forall h = 0,...,t.
$$

\noindent Similarly, we also define the \textit{local residual treatment effect} as:

$$
\Delta_{R}(t)=\int  \mathbb{E}[Y\supone_t- Y\supzero_t| \bS\supone_t = \bS\supzero_t=\boldsymbol{s}_t]dF_{\bS\supzero_t}(\boldsymbol{s}_t).
$$

\noindent Note here that we only conditioned on the surrogate up to time $t$; this is due to the nature of the process we investigate, since for a time series, we assume that causality has an exclusive direction: the past affects the future, and not the other way around. Thus, given the present and past surrogate values, the future surrogate values have no causal effect on the present outcome. With these definitions, we now define our main quantity of interest: the \textit{proportion of treatment effect explained by the surrogate} (PTE) in the longitudinal setting:

$$
\PTE = \frac{\Delta_{\text{area}} - \Delta_{R,\text{area}}}{\Delta_{\text{area}}} = 1 - \frac{\Delta_{R,\text{area}}}{\Delta_{\text{area}}}.
$$

\noindent We also define the Local PTE (LPTE):

$$
\LPTE(t) = 1 - \frac{\Delta_{R}(t)}{\Delta(t)}.
$$

\noindent and the Cumulative PTE (CPTE):

$$
\CPTE(t) = 1 - \frac{\sum_{h=0}^{t}\Delta_{R}(h)}{\sum_{h=0}^{t}\Delta(h)}.
$$

\noindent Naturally, $\PTE = \CPTE(T)$ and we can express $CPTE(t)$ as:

$$
\CPTE(t) = 1-\frac{\sum_{h=0}^{t}\Delta(h)(1-\LPTE(h))}{\sum_{h=0}^{t}\Delta(h)} =\sum_{h=0}^{t} \frac{\Delta_{h}}{\sum_{i=0}^{t}\Delta_i}\LPTE(h).
$$

\noindent That is, the CPTE can be interpreted as the weighted average of the LPTE, with each time having weight proportional to the treatment effect at that particular time. This also shows that the CPTE is constant over time if, and only if, the LPTE is also constant over time. In practice, a PTE estimate (or the lower bound of the confidence interval) greater than 0.5 or 0.75 is often considered a ``strong" surrogate marker \citep{lin1997estimating}. Importantly, the identifiability of the PTE definitions above rely on the following condition assumed to hold throughout:

\begin{enumerate}
    \item[\textbf{(D4)}]  Identifiability:
    $$
\int \mathbb{E}\left[Y_t^{(g)}\middle|\bS\supone_t=\bS\supzero_t=\boldsymbol{s}\right]dF_{\bS\supzero}(\boldsymbol{s}) = \int \mathbb{E}\left[Y_t^{(g)}\middle|\bS^{(g)}_t=\boldsymbol{s}\right]dF_{\bS\supzero}(\boldsymbol{s}),
    $$
    for all times $t$ and both treatment arms $g$.
\end{enumerate}

\noindent Although this assumption is not unique to our setting, its extension to the longitudinal setting requires conditioning on the full trajectory.


Our proposed definitions highlight a particular complexity of evaluating a surrogate in this setting: it is possible for a potential surrogate, for example a biomarker, to be a strong surrogate at specific times, but not for others, or possibly be a strong up to a certain point, but not afterwards. Given these complexities, we will additionally propose a hypothesis test to formally test whether there is evidence that the PTE changes over time in Section \ref{subsec:hetero_test}.

Lastly, while the identifiability condition \textbf{(D4)} is sufficient for estimating the residual treatment effect, by itself this condition does not guarantee that the PTE can be interpreted as a proportion in the traditional sense; in \ref{app:non-neg} we discuss additional conditions that would ensure the PTE is between 0 and 1.

\subsection{Model Conditions}\label{subsec:model_assumptions}
Thus far, we have defined a causal framework for surrogate evaluation via the PTE when both the outcome and surrogate are longitudinal. In Section \ref{sec:estimating} we will provide a review of state-space models (SSMs) and propose a SSM approach to estimate the PTE. However, discussion of some preliminary details is warranted. In particular, regardless of the exact chosen modeling strategy, there are certain conditions that are needed to ensure that the PTE quantities are estimable. These conditions, which we detail here, will be needed for any approach that involves linear models. We consider two models: one for $Y_{i,t}\mid \bY_{i,t-1},G_i$ and another for $Y_{i,t}\mid \bY_{i,t-1},\bS_{i,t},G_i$. We refer to the first model as the \textit{marginal model}, and to the second as the \textit{conditional model}. We purposefully keep this quite general in this section, and denote the estimators of $\Delta_R(t)$ and $\Delta(t)$ that follow from fitting these specified linear models as $\widehat{\Delta}_R(t)$ and $\widehat{\Delta}(t)$. We will explicitly express these estimators under a particular SSM choice for these two models in Section \ref{sec:estimating}.

Assuming this two-model framework, we define for each model $m \in \{\text{marginal}, \text{conditional}\}$, the stacked working mean model as:
$$
Y_m = X_m \eta_m + \varepsilon_m,
$$
where $Y_m$ is the stacked outcome vector, $X_m$ is the corresponding fixed-effect design matrix, and $\eta_m$ is the vector of fixed-effect parameters. Let $A_m$ denote the working precision matrix used by the Generalized Least Squares (GLS) estimator, which may be induced by an ordinary GLS, Linear Mixed Model (LMM), or SSM procedure. The corresponding estimator is:
$$
\widehat{\eta}_m = (X_m^\top A_m X_m)^{-1}X_m^\top A_mY_m . 
$$

\noindent Let $\eta_m^*$ denote the pseudo-true value of $\eta_m$ under the working mean structure, and define:
$\varepsilon_m^* = Y_m - X_m\eta_m^*
$. We assume the following model conditions:
{
\begin{enumerate}
    \item[\textbf{(C1)}]  \textbf{Fixed duration and equally spaced measurements:} The primary outcome $Y_{i,t}$ and surrogate $S_{i,t}$ are measured at discrete, equally spaced time points $t = 0,\dots,T$ for each subject $i$.
    \item[\textbf{(C2)}]  \textbf{Finite moments:} 
    $
    \mathbb{E}[Y_{i,t}]<\infty \text{ and }\mathbb{E}[Y_{i,t}^2]<\infty, \text{for all } i,t.
    $
    \item[\textbf{(C3)}] \textbf{Common marginal distribution and weak cross-subject dependence:} For each working model $m$, the subject-level trajectories have the same marginal distribution, and $
    \frac{1}{N}X_m^\top A_m\varepsilon_m^\ast \overset{p}{\longrightarrow}0 .$
    
    \item[\textbf{(C4)}] \textbf{Nondegenerate GLS information:} For each working model $m$, the working precision matrix \(A_m\) is symmetric positive definite with probability tending to one. In addition, there exist constants $0<c<C<\infty$ such that
    
    $$
    P\left[ c \leq \lambda_{\min}\!\left\{ \frac{1}{N}X_m^\top A_mX_m \right\} \leq \lambda_{\max}\!\left\{ \frac{1}{N}X_m^\top A_mX_m \right\} \leq C \right] \longrightarrow 1,
    $$
    where $\lambda_{\min}(M)$ and $\lambda_{\max}(M)$ are the smallest and largest eigen values of $M$, respectively.

        \item[\textbf{(C5)}] \textbf{Square-integrable potential outcome regression:} 
$$
 \int \mathbb{E}\left[Y_t^{(1)}\middle|\bS^{(1)}_t=\boldsymbol{s}\right]^2dF_{\bS\supzero}(\boldsymbol{s})<\infty
$$

    \item[\textbf{(C6)}] \textbf{Control-distribution unbiasedness:} 
Let: $q_{1,t}(s)=\mathbb{E}[Y_{i,t}\mid G_i=1,S_{i,t}=s]$
denote the true treated conditional mean, and let $q^*_{1,t}(s)$ denote the
pseudo-true treated working mean of the chosen fitting procedure. We require:
$$
\int
\{q^*_{1,t}(s)-q_{1,t}(s)\}
\,dF_{S_t\mid G=0}(s)=0,
\qquad t=0,\ldots,T.
$$
    \end{enumerate}

\noindent We discuss condition \textbf{(C1)-(C5)} in detail in \ref{app:conditions}, and focus here on  Condition \textbf{(C6)}.  Condition \textbf{(C6)} can be satisfied in multiple ways. One approach is to use an outcome-regression model with the surrogate trajectory as a covariate. The working mean need not be correctly specified pointwise; rather, its approximation error must average to zero under the control-group surrogate distribution. A correctly specified model, for which $q^*_{1,t}(s)=q_{1,t}(s)$, trivially satisfies this condition. Alternatively, under a bias-correction approach proposed in \ref{app:pseudo}, \textbf{(C6)} can hold even when the working conditional mean is misspecified. In that case, the treated working mean is fit using weights that target the control-group surrogate distribution, so that the corresponding projection residual has mean zero under $F_{S_t\mid G=0}$. Lastly, randomization is sufficient to guarantee consistency for the marginal treatment effect, even under working-mean misspecification, so no additional conditions are necessary in that regard. It is also worth distinguishing between two forms of misspecification: general misspecification of the working mean, discussed above, and omission of lingering effects, which will be discussed in Section~\ref{subsec:lingering}. 


An additional modeling consideration worth clarifying is the use of a separate marginal and conditional model. Although, in principle, the conditional model implies the marginal model, suggesting that specifying both may be redundant, we adopt this approach for robustness. The conditional model, which must capture lingering effects, is inherently more susceptible to misspecification. We will show that, under appropriate conditions, misspecification of the conditional model leads to underestimation of the PTE, resulting in a conservative error. However, this guarantee relies on the estimator for the total treatment effect being consistent, which holds under randomization. If inference were based solely on the conditional model, it would no longer be possible to bound the bias induced by the omission of lingering effects. We discuss these issues in detail in Section~\ref{subsec:lingering}.

Conditions \textbf{(C1)--(C6)} guarantee consistency specifically for the point estimator of the fixed effects, that is, the ones that are shared among patients. Importantly,  $\Delta_{R}(t)$ and  $\Delta(t)$ are among the fixed effects, and thus, the Kalman filter, which we will use, (and Generalized Least Squares (GLS) in general) provides consistent estimator for $\Delta_{R}(t)$ and  $\Delta(t)$. We formally state the consistency of the estimators in Theorem \ref{theo:pte}.

\vspace*{5mm}

\begin{theorem}[Consistency of estimators of the longitudinal treatment effects]
\label{theo:pte}
Let $N_g$ be the number of people under treatment $g$, with $N=N_0+N_1$ denoting the total sample, and  $\underset{N\rightarrow \infty}{\lim \sup}\,  N_0/(N_0 + N_1) <1$ and ~$\underset{N\rightarrow \infty}{\lim \inf} ~N_0/(N_0 + N_1) >0$.  Under  \textbf{(C1)--(C6)}, 
$$
\widehat{\Delta}_{R}(t)\underset{N\rightarrow\infty}{\xrightarrow{p}}\widetilde{\Delta}_{R}(t),
\qquad
\widehat{\Delta}(t)\underset{N\rightarrow\infty}{\xrightarrow{p}}\widetilde{\Delta}(t),
$$
where $\widetilde{\Delta}_{R}(t)$ and $\widetilde{\Delta}(t)$ are the pseudo-true limits defined by the working mean model structure, detailed in \ref{app:pseudo}. Under \textbf{(D4)}, $$\widetilde{\Delta}_{R}(t)=\Delta_{R}(t) \mbox{~~and~~} \widetilde{\Delta}(t)=\Delta(t).$$ Thus, under \textbf{(D4)} and \textbf{(C1)--(C6)}, if  $ \sum_{t=0}^T\Delta(t)\neq0$ then 
$$
1-\frac{\sum_{t=0}^{T}\widehat{\Delta}_{R}(t)}{\sum_{t=0}^{T}\widehat{\Delta}(t)}\xrightarrow{p}\CPTE(T) = PTE
$$
\end{theorem}

\vspace*{5mm}

\noindent The following proof sketch outlines the main arguments. The claims in Theorem \ref{theo:pte} follow from consistency of the GLS/Kalman estimators for the working mean model, together with continuous mapping arguments for the resulting plug-in estimators. Specifically, under \textbf{(C1)--(C6)}, the fixed–effects are consistent (and asymptotically normal) for their pseudo–true limits. Then, under \textbf{(D4)}, the pseudo–true limits coincide with the population targets. A further application of the continuous mapping theorem then yields consistency of the plug-in estimator of the PTE. 

Regarding definability of the PTE, it suffices that $ \sum_{t=0}^T\Delta(t)\neq0$; otherwise, the ratio is undefined. Even if this condition fails, the components $\Delta_{R}(t)$ and $\Delta(t)$ remain consistently estimable, which allows one to assess definability empirically by testing $H_0:\sum_{t=0}^T\Delta(t)=0$ or examining confidence intervals for $\sum_{t=0}^T\Delta(t)$. Lastly, although consistency holds under \textbf{(C1)--(C6)}, the finite-sample performance of the estimator may deteriorate when $Y_t$ is concentrated near the boundary of its support, in which case a substantially large sample size may be required.

For uncertainty quantification, multiple approaches can be used. In Section \ref{subsec:bootstrapping} we discuss a nonparametric bootstrap approach for SSMs. This approach requires an additional set of conditions to provide valid uncertainty for our estimands.

\begin{enumerate}
    \item[\textbf{(B1)}]  \textbf{Subject-level i.i.d. units:} The individual trajectories $(\mathbf{Y}_i,\mathbf{S}_i,G_i)$ are independent and identically distributed across $i$. That is, the joint distribution of the outcome and surrogate  trajectories, as well as treatment assignment, is the same for all individuals, and individuals are mutually independent.
    \item[\textbf{(B2)}]  \textbf{PTE is well-defined:} Beyond \textbf{(C1)} and \textbf{(C2)}, the total treatment effect satisfies $\sum_{t=0}^{T}\Delta(t) \ge \epsilon$ for some $\epsilon>0$, so that the PTE is well-defined and nondegenerate.
    \end{enumerate}

\noindent If \textbf{(B2)} is violated, the proposed inference for the residual and marginal treatment effects remains valid, but the PTE is undefined due to a degenerate total treatment effect.
}

\subsection{Lingering Effects}\label{subsec:lingering}

Thus far, the problem of surrogate evaluation when both the outcome and the surrogate marker are longitudinal may not seem particularly different from an endpoint analysis or simply a non-longitudinal problem with repeated measurements for the same subjects. Here, we discuss the main challenge that makes our setting unique: the lingering effect of the surrogate on the outcome. 

Ideally, one would like to account for all causal pathways linking the treatment to the outcome. For each time point $t$, this would require incorporating the treatment effect captured by every prior measurement of the surrogate marker. In practice, however, it is often infeasible to use the entire surrogate trajectory at each time point. This may be due to the substantial computational burden of specifying a model that can accommodate all relevant interactions between past surrogate values and the current outcome, or to the loss of statistical power that can arise from an overly flexible specification, which may yield unstable or uninformative estimates. When one has the means to model the full surrogate trajectory, and a sufficiently large sample size to obtain meaningful estimates, this is the ideal approach. When this is not feasible, we discuss the consequences of omitting part of the surrogate trajectory. The goal of this section is therefore two-fold: 1) to discourage the omission of surrogate values unless strictly necessary, by showing that doing so biases the PTE estimator, and 2)  providing insight into the consequences of such omissions in settings where they cannot be avoided, specifically showing that the resulting estimator does not become entirely invalid.

In practice, we argue that older surrogate values are likely to be less informative about the current outcome than more recent measurements. Accordingly, if simplification of the surrogate trajectory is required, older values would typically be omitted first. In the extreme scenario where the entire surrogate history is discarded and only the current surrogate value is used, the resulting estimator of the treatment effect component within the residual treatment effect converges to
$$
  \mathbb{E}\!\left[Y_t^{(1)}-Y_t^{(0)} \,\middle|\, S_t^{(1)} = S_t^{(0)}\right]
  \qquad\text{instead of}\qquad
  \mathbb{E}\!\left[Y_t^{(1)}-Y_t^{(0)} \,\middle|\, \bS_t^{(1)} = \bS_t^{(0)}\right].
$$
Although this estimand is not useless, it is not our target.  Since some of the total treatment effect may be captured by $S_{t-k}$ for some $k>0$; omitting those values risks incorrectly attributing some of that effect as part of the residual effect, and thus biasing the PTE estimator. Alternatively, an arguably more defensible approach would be one that includes the last $k$ surrogate values, in which case the estimator would converge to
$$
  \mathbb{E}\!\left[
    Y_t^{(1)}-Y_t^{(0)}
    \,\middle|\,
    S_t^{(1)} = S_t^{(0)},\dots,
    S_{t-k}^{(1)} = S_{t-k}^{(0)}
  \right],
$$
which would be closer to, though still not the same as, the desired
$\mathbb{E}\!\left[Y_t^{(1)}-Y_t^{(0)} \middle| \bS_t^{(1)} = \bS_t^{(1)}\right]$.

At this point, we suppose that we are in a situation where it is infeasible to include the entire surrogate trajectory, but yet we have also shown that if we include only the most recent $k$ values, we end up with a biased estimator, which leaves us in a difficult spot. However, if we can show that including only the most recent $k$ values (omitting older values) result in a smaller PTE estimate, compared to the truth, then this omission would make us more likely to reject the surrogate i.e., conclude that it is not a strong surrogate. Thus, making a decision based on the truncated PTE estimator would be a \textit{conservative} approach. 

To this end, suppose that instead of estimating: 

$$\Delta_{R,\text{Area}} =  \sum_{t=0}^{T}\int  \mathbb{E}\left[Y\supone_t- Y\supzero_t\middle| S\supone_h = S\supzero_h={s}_h, h=0,\cdots,t\right]dF_{\bS\supzero_t}(\boldsymbol{s}_t),$$
we estimate:
$$
\Delta^*_{R,\text{Area}} =  \sum_{t=0}^{T}\int  \mathbb{E}\left[Y\supone_t- Y\supzero_t\middle| S\supone_h = S\supzero_h=s_h, h = t - k,\cdots,t\right]dF_{\bS\supzero_t}(\boldsymbol{s}_t),
$$
for some integer $0\le k \le t$ and where $F_{\bS\supzero_t}$ represents the density for $\bS\supzero_t$, the surrogate trajectory up to time $t$.
Note that $\Delta_{R,\text{Area}}^{*}$ is the residual treatment effect considering only the most recent $k$ values of the surrogate, and not the whole trajectory. Let $\PTE^{*} = 1-\Delta^{*}_{R,\text{Area}}/\Delta_{\text{area}}$, which we refer to as the truncated PTE. In \ref{app:lingering}, we present formal conditions to guarantee that $\PTE^{*} \le \PTE$.


Although unbiased estimation is ideal, conservativeness is a desirable property in surrogate evaluation. Overestimation of the PTE may lead to incorrectly deeming a surrogate to be strong, which in turn could justify its use to test for a treatment effect in future studies. In contrast, underestimation results in a more cautious assessment and mitigates the risk of endorsing an invalid surrogate.

\section{A State-Space approach for estimating the PTE}\label{sec:estimating}
In this section, we first briefly review state-space models and then describe our proposed estimation procedure, uncertainty quantification, and temporal homogeneity test.

{
\subsection{State-space Models}\label{subsec:state-space}

State-space models have enjoyed considerable success since their conception, with widespread applications across fields such as econometrics, ecology, and epidemiology \citep{Alves2025Econometrics,Prashad2025Epidemiology}. However, to our knowledge, use of such models for surrogate evaluation has not yet been considered. A State-Space Model (SSM), or Dynamic Linear Model (DLM), is defined by the following set of equations:
\vspace*{-3mm}
\begin{align}
    Y_t &= F_t^{\top}\btheta_t+\bepsilon_t, \quad\bepsilon_t \sim [\boldsymbol{0},V_t]\label{eq.1}\\ 
    \btheta_t &= G_t\btheta_{t-1}+\bomega_t,\quad \bomega_t \sim [\boldsymbol{0},W_t] \label{eq.2}
\end{align} \vspace*{-14mm}

\noindent where $F_t$ and $G_t$ are known matrices, called the design matrix and the evolution matrix, respectively. We use the notation $X\sim [m,s^2]$ to denote that $X$ is a random variable with mean $m$ and variance $s^{2}$ (no assumptions on the specific form of the distribution). Equation (\ref{eq.1}) is referred to as the \textit{Observation equation} or \textit{Observational model} and can be generalized  for non-Gaussian models \citep{WestHarrMigon,Alves-kparametric}. Equation (\ref{eq.2}) is referred to as the \textit{System equation} or \textit{Evolution equation} and can be expanded to include unknown components in the matrix $G_t$ to account for non-linear temporal dynamics \citep[see][Section 13.2]{West-DLM}. The design matrix $F_t$ is analogous to the design matrix $X$ in the context of linear regression and includes structural components of the model (trend, seasonality, etc.) and covariates. For our purpose in this paper, $F_t$ will contain (at least) an indicator of the treatment group, such that one of the components of $\btheta_t$ will represent the effect of the  treatment at time $t$.

Beyond their highly flexible and intuitive interpretation, state-space models are popular because it is straightforward to obtain estimates of $\btheta_t$. In particular, the Kalman filter and smoother algorithms \citep{Kalman_filter_origins} provide sequential, highly efficient procedures for computing point and interval estimates of $\btheta_t$. As noted by \cite{Arulampalam-particlefilter}, the estimator obtained by the Kalman filter is a particular case of the Generalized Least Squares (GLS) estimator. Therefore, the Kalman filter estimator is valid, in the sense that it provides a consistent estimator as long as conditions \textbf{(C1)--(C6)} hold. In \ref{app:ssmolslmm}, we discuss how the SSM approach is similar to and differs from ordinary least squares (OLS) and a linear mixed model (LMM), which are also particular cases of the GLS. Furthermore, if $V_t$ and $W_t$ are correctly specified, the estimator is also efficient, that is, it has the smallest variance within the class of linear estimators for $\btheta_t$. Naturally, it is unreasonable to assume that $V_t$ and $W_t$ are known, but by using consistent estimators of these quantities, one  can achieve a close-to-efficient estimator.
}

\subsection{Estimating the PTE}\label{subsec:estimating}

Next, we propose a specific SSM to estimate the PTE using observed data. The observed data include: $Y_{i,t}$, the outcome for the individual $i$ measured at time $t$, $S_{i,t}$, the surrogate of the individual $i$ measured at time $t$, and $G_i$, which  indicates if individual $i$ received the treatment, such that $G_i=1$ if, and only if, the $i$th individual received the treatment. We specify the following models:

\begin{enumerate}
    \item Conditional model:
    $$
    \begin{aligned}
    Y_{i,t} &= \mu_{t}+m_{i,t}+f^{(G_i)}_t(S_{i,t})+\delta_{1,t} G_{i} + \epsilon_{i,t},\quad \epsilon_{i,t} \sim \left[0,V_t^{(G_i)}\right]\\
    \mu_{t} &= \mu_{t-1} + \omega_t,\quad \omega_{t} \sim [0,W_{\mu,t}]\\
    m_{i,t} &= m_{i,t-1} + w_{i,t},\quad w_{i,t} \sim [0,W_{i,t}]
    \end{aligned}
    $$
    \item Marginal model:
    $$
    \begin{aligned}
    Y_{i,t} &= \nu_{t}+n_{i,t}+\delta_{2,t} G_{i} + \epsilon'_{i,t},\quad \epsilon'_{i,t} \sim \left[0,V_t^{'(G_i)}\right]\\
    \nu_{t} &= \nu_{t-1} + \omega'_{t},\quad \omega'_{t} \sim [0,W'_{\nu,t}]\\
    n_{i,t} &= n_{i,t-1} + w'_{i,t},\quad w'_{i,t} \sim [0,W'_{i,t}]\end{aligned}
    $$
\end{enumerate}

\noindent where $f^{(g)}_t$ is a nonlinear, time- and treatment-dependent surrogate effect; $W_{\mu,t}$, $W_{i,t}$, $W'_{\nu,t}$ and $W'_{i,t}$ are the evolution covariance matrices; and $V_t(g)$ and $V'_t(g)$ are observational covariance matrices, for $g=0,1$. The first line in each model corresponds to the observational equation (Equation \ref{eq.1}), while the remaining lines correspond to the evolution equation (Equation \ref{eq.2}) described in Section \ref{subsec:state-space}, such that $G_t$ is the identity matrix, and $F_t$ is a column vector of $1$’s and $0$’s indicating treatment.

Of course, $f^{(g)}_t$ is unknown and must be estimated; in Section \ref{sec:simul}, we  use splines and express $f_t^{(g)}$ as a linear function of known basis (e.g., indicator) functions of the surrogate. { To avoid over-parametrization, we specify the evolution covariance matrices $W_{\mu,t}$, $W_{i,t}$, $W'_{\nu,t}$ and $W'_{i,t}$ using a discount strategy, detailed in \ref{app:discount}. Importantly, neither model assumes  that the treatment has the same effect at each time point, as can be seen by the parameters $\delta_{1,t}$ and $\delta_{2,t}$ which depend on $t$.  Lastly, note that $\mu_t$, $m_{i,t}$, and $f^{(g)}_t$ are identifiable only up to their sum, $\mu_t + m_{i,t} + f^{(g)}_t(s)$. This is not a problem for our purpose, since the treatment effects remain identifiable and the Kalman filter’s regularization penalizes the latent components, allowing them to be estimated. Under the marginal model, it follows that:
$$
\mathbb{E}[Y_t^{(1)} - Y_{t}^{(0)}] = \delta_{2,t} \Rightarrow \Delta_{\text{area}} = \sum_{t=0}^{T}\delta_{2,t}
$$

\noindent Under the conditional model and \textbf{(D4)}, it follows that:
$$\begin{aligned}
   \Delta_{R}(t) &= \int \left \{ \mathbb{E}[Y_{t}^{(1)}|\bS_{t}^{(1)}=\bs_t] - \mathbb{E}[Y_{t}^{(0)}|\bS_{t}^{(0)}=\bs_t] \right \} dF_{\bS\supzero_t}\\
    &= \delta_{1,t}+\int \left \{ f_{t}^{(1)}(\bs_t) - f_{t}^{(0)}(\bs_t) \right \}dF_{\bS\supzero_t}\\
&\Rightarrow \Delta_{R,\text{area}} = \sum_{t=0}^{T}\delta_{1,t} + \sum_{t=0}^{T}\int \left \{ f_{t}^{(1)}(\bs_t) - f_{t}^{(0)}(\bs_t)\right \} dF_{\bS_{t}\supzero},
\end{aligned}
$$

\noindent where the integral can be calculated using the empirical distribution of the surrogate in the control group. With this structure, the conditional model provides a consistent estimator for the treatment effect, and thus Theorem \ref{theo:pte} holds for this model.

It follows that the estimator for the PTE is:

$$
\widehat{\PTE} = 1-
\frac{\sum_{t=0}^{T}\widehat\delta_{1,t}\;+\;\sum_{t=0}^{T}\frac{1}{n_0}\sum_{i:G_i=0}\Big(\widehat f_t^{(1)}(S_{i,t})-\widehat f_t^{(0)}(S_{i,t})\Big)}
{\sum_{t=0}^{T}\widehat\delta_{2,t}}.
$$

\noindent where $\widehat{\delta}_1$, $\widehat{\delta}_2$, $\widehat{f}_{t}^{(1)}(\bS_{i,t})$ and $\widehat{f}_{t}^{(1)}(\bS_{i,t})$ are obtained using the Kalman Filter and Smoother algorithms. We construct similar estimators for the local and cumulative PTE, provided in \ref{app:estimate}. Both models can be expanded to include baseline covariates and/or interaction terms, and the conditional model can be expanded to include additional surrogate marker measurements. We work with these models here solely for notational convenience for the subsequent expressions. In our numerical studies, we indeed use expanded versions of these models as described in Sections \ref{sec:simul} and \ref{sec:app}.

\subsection{Uncertainty Quantification for Gaussian SSMs}\label{subsec:bootstrapping}

In the prior section, we proposed a method to estimate the PTE. In practice, there is interest in not only the PTE point estimate, but also the corresponding confidence interval. Though this is generally the case in statistics, it is especially important in the surrogate evaluation setting because decisions about surrogate validity, and whether to use the surrogate as a replacement of the primary outcome in a future trial, are typically based on the lower bound of the confidence interval being greater than some pre-specified value such as 0.75. We propose an approach to estimate the standard error of the PTE estimate and corresponding confidence interval by introducing a nonparamtric bootstrap strategy for SSMs in longitudinal studies; specific details describing this proposed procedure are provided in \ref{app:boot}. Notably, a naïve bootstrap with $B$ replications, resampling entire individual trajectories each time, costs $\mathcal{O}(BNT)$, since a single fit costs $\mathcal{O}(NT)$. Our proposed approach reduces the total cost to $\mathcal{O}(NT)+\mathcal{O}(BN)$, yielding a much more friendly scaling. We use these resulting subject-level bootstrap estimates to calculate the standard error of our proposed estimators and percentile-based confidence intervals for the true quantities. We implement and investigate the performance of this proposed procedure in Section \ref{sec:simul}.

\subsection{Testing Temporal Homogeneity of the PTE}\label{subsec:hetero_test}

As was previously mentioned, it is possible that the PTE may change over time and such knowledge would be useful in practice because it may change decisions about whether and/or when to use the surrogate as a replacement of the primary outcome in future studies. Thus, in this section, we propose a test for temporal homogeneity of the PTE using our proposed estimator and bootstrap standard error estimates. We leverage prior work in the non-temporal setting which proposed an omnibus test for PTE heterogeneity with respect to a baseline covariate, described in \cite{parast2021hetero}. In particular, \cite{parast2021hetero} focused on testing whether the PTE, estimated as a function of a person-level baseline covariate such as age, varied with respect to this covariate. Specific to our setting, we first note that the CPTE is constant if, and only if, the LPTE is constant, and thus,  we focus our test construction on the LPTE, which is composed of $\Delta_{R}(t)$ and $\Delta(t)$. If the LPTE is constant, i.e., equal to some $\tau$, it follow that:

$$
1-\frac{\Delta_{R}(t)}{\Delta(t)} = \tau, \forall t.
$$

\noindent which is equivalent to  $\Delta_{R}(t)- (1-\tau)\Delta(t)=0, \forall t$. This motivates us to define $\Delta_{\text{diff}}(t) \equiv \Delta_{R}(t)- (1-\tau)\Delta(t)$ such that testing whether the PTE is constant over time is equivalent to testing if $\Delta_{\text{diff}}(t)=0$. Since $\Delta_{\text{diff}}(t)$ is a linear transformation of $\Delta_{R}(t)$ and $\Delta(t)$, it is trivial to derive an estimator for the former based on the estimator for the latter. With a slight abuse of notation, we define $\widehat\Delta_{\text{diff}}=\left(\widehat\Delta_{\text{diff}}(1),\dots,\widehat\Delta_{\text{diff}}(T)\right)'$, where $\widehat\Delta_{\text{diff}}(t)  =  \widehat\Delta_{R}(t)- (1-\widehat{\tau})\widehat\Delta(t)$ and let $\widehat{\tau}$ denote our estimate of the global PTE i.e., the CPTE estimate at time $T$. We aim to test the following:
$$
\begin{aligned}
H_0: \Delta_{\text{diff}}(t)={0},\forall t,\quad\quad \quad
H_1: \Delta_{\text{diff}}(t)\neq {0} ,\text{ for some }t 
\end{aligned}
$$

This statement of hypotheses makes a straightforward Wald-based test a natural approach to consider. However, it turns out that such a test can perform quite poorly; we illustrate and discuss this further in Section \ref{sec:simul}. Thus, we do not propose a Wald-based test and instead propose the following test statistic:

$$
\mathcal{T}= \max_{t} \left|\frac{\widehat{\Delta}_{\text{Diff}}(t)}{\widehat{\sigma}(t)}\right|,
$$
where $\widehat{\sigma}^2(t)$ is the estimator of the variance of $\widehat{\Delta}_{\text{Diff}}(t)$, obtained by our proposed bootstrap procedure. This statistic parallels the construction of the omnibus test in \cite{parast2021hetero}, but adapted to the case where we aim to test temporal homogeneity, i.e. homogeneity in a discrete range of values, specifically, the time index. We refer to our proposed test as the Maximum Standardized Deviation (MSD) test. To obtain the critical value of this test, we sample from the asymptotic distribution of $\widehat{\Delta}_{R}(t)$ and $\widehat{\Delta}(t)$ under $H_0$, and then compute the critical value as the $c_\alpha$ as the $100(1-\alpha)th$ percentile of the sampled distribution. We formally state the properties of this test in Theorem \ref{thm:test-consistency}.

\begin{theorem}
\label{thm:test-consistency}

Under conditions \textbf{(C1)}--\textbf{(C6)} and \textbf{(B1)}--\textbf{(B2)}, it holds that:
\begin{enumerate}
\item[(i)] The test based on $\mathcal T$ controls the type I error at the nominal significance level asymptotically.
\item[(ii)] The test is consistent against fixed alternatives: as $N \to \infty$, $\mathcal T \xrightarrow{p} \infty$ under $H_1$, and the type II error converges to $0$.
\end{enumerate}
\end{theorem}

\noindent The following proof sketch outlines the main arguments. Under \textbf{(C1)}--\textbf{(C6)} and \textbf{(B1)}--\textbf{(B2)}, the vector of estimators $\left\{\widehat{\Delta}_R(t),\widehat{\Delta}(t):t=0,\ldots,T\right\}$ can be written as a smooth functional of the empirical distribution of the subject-level trajectories. {
As discussed in \ref{app:boot}, under \textbf{(B1)} and \textbf{(B2)} the subject-level bootstrap consistently approximates the joint sampling distribution of $\left\{\widehat{\Delta}_R(t),\widehat{\Delta}(t)\right\}$. Since $\widehat{\Delta}_{\mathrm{Diff}}$ is a continuously differentiable transformation of $\left\{\widehat{\Delta}_R(t),\widehat{\Delta}(t):t=0,\ldots,T\right\}$, the  bootstrap also consistently approximates the sampling distribution of $\widehat{\Delta}_{\mathrm{Diff}}$.
} This establishes the asymptotic validity of the bootstrap critical value and, consequently, the type I error control in part (i). For part (ii), under any fixed alternative, there exists at least one time point $t^{*}$ such that $|\Delta_{\mathrm{Diff}}(t^{*})|>0$. By consistency of $\widehat{\Delta}_{\mathrm{Diff}}(t)$, $\widehat{\Delta}_{\mathrm{Diff}}(t^*) \xrightarrow{p} \Delta_{\mathrm{Diff}}(t^*)$, while the bootstrap standard error satisfies $\widehat{\sigma}(t^{*})=O_p(N^{-1/2}).$ Therefore:
$$
\left| \frac{\widehat{\Delta}_{\mathrm{Diff}}(t^{*})}{\widehat{\sigma}(t^{*})} \right| \xrightarrow{p} \infty,
$$ 
and since $\mathcal{T}= \max_{t} \left|\frac{\widehat{\Delta}_{\text{Diff}}(t)}{\widehat{\sigma}(t)}\right|\ge \left| \frac{\widehat{\Delta}_{\mathrm{Diff}}(t^{*})}{\widehat{\sigma}(t^{*})} \right| ,$ it follows that  $\mathcal T \xrightarrow{p}\infty$. Thus, the probability of rejecting $H_0$ converges to one, and the type II error converges to zero.
In Section \ref{sec:simul}, we implement and investigate the performance of our proposed test, both under the null and various alternatives.

\vspace*{-4mm}
\section{Simulation Study}\label{sec:simul}

In this section, we assess the performance and robustness of our methods in finite samples via a simulation study. Our primary goals were (1) to assess bias and coverage of the proposed PTE estimator and corresponding confidence interval; (2) to investigate the performance of a test for surrogate validity based on the lower bound of our constructed PTE confidence interval; and (3) to investigate the performance of the proposed test for temporal homogeneity. For goals (1) and (2), we examined 15 different simulation settings with varying surrogate strength (from PTE $=$ 0.25 to 1.0) and varying treatment effect trajectories (monotone, parabole, random walk), each of which was investigated across 6 different sample sizes (50 to 300). For goal (3), we examined 5 main settings which varied with respect to the treatment effect, residual treatment effect, and PTE trajectories over time, each of which was investigated across 6 different sample sizes (50 to 300) and 5 trial lengths (1 year to 10 years).  The data generation details for all settings are provided in \ref{app:sims}. For all settings, the over-arching theme was that data were purposefully generated such that the specified models in our estimation procedure do \textit{not} exactly hold. For example, the distributions of the observational and evolution errors were purposefully non-symmetric, heavy-tailed, and involve nonlinear temporal dynamics. Thus, we implicitly assessed robustness of our proposed methods to temporal misspecification and error non-normality. Throughout, we compare our proposed methods to alternatives that would be logical to consider in practice. All results are summarized over 2,000 simulation iterations for the first surrogate evaluation study and 3,000 iterations for the temporal homogeneity study, with 2,000 bootstrap iterations used for uncertainty quantification.

\vspace*{-4mm}
\subsection{Results: Surrogate Evaluation}
First, we assessed the finite sample performance of the proposed PTE estimator and corresponding confidence interval, and of a test for surrogate validity based on the lower bound of our constructed PTE confidence interval. As described in \ref{app:sims}, the simulation settings were such that the treatment effect changed over time, but the PTE was homogeneous over time (in the following section, the PTE will not be homogeneous over time). We compared our estimator to the following: a naive OLS estimator, a Linear Mixed Model (LMM), and a Generalized Estimating Equation (GEE) estimator. When relevant, we also compared our approach with a method that considers the simple change in treatment effect from the beginning to the end of the study, using the corresponding change in the surrogate; we refer to this approach as \textit{Diff}. This approach is only meaningful when the treatment effect is monotonic, and it fails to capture several insights that our framework can reveal. Nevertheless, we include this comparison whenever possible to demonstrate that, even under these conditions favorable to the \textit{Diff} approach, our proposed framework performs better. All other alternative approaches remain within our framework, but they differ in how they estimate the treatment effect at each time point. The purpose of these comparisons is to show that a model designed to explicitly capture the temporal structure of the data outperforms standard methods that either assume independence over time or rely on more generic dependence assumptions.

With respect to the PTE estimator, Figure \ref{fig:simul3_bias} shows the average bias (i.e., $|\PTE-\widehat{\PTE}|$) of the proposed and comparison estimators across all settings and demonstrates that our proposed approach performs well, with small bias, generally outperforming the comparison methods. In \ref{app:sims2}, we demonstrate (1) coverage close to the nominal level of 0.95 for the proposed method in all settings, and over-coverage for the LMM and GEE in some settings, and (2) confidence interval widths that are smallest for our proposed method. 

\if\figureend1
\begin{figure}
    \centering
    \includegraphics[scale=1]{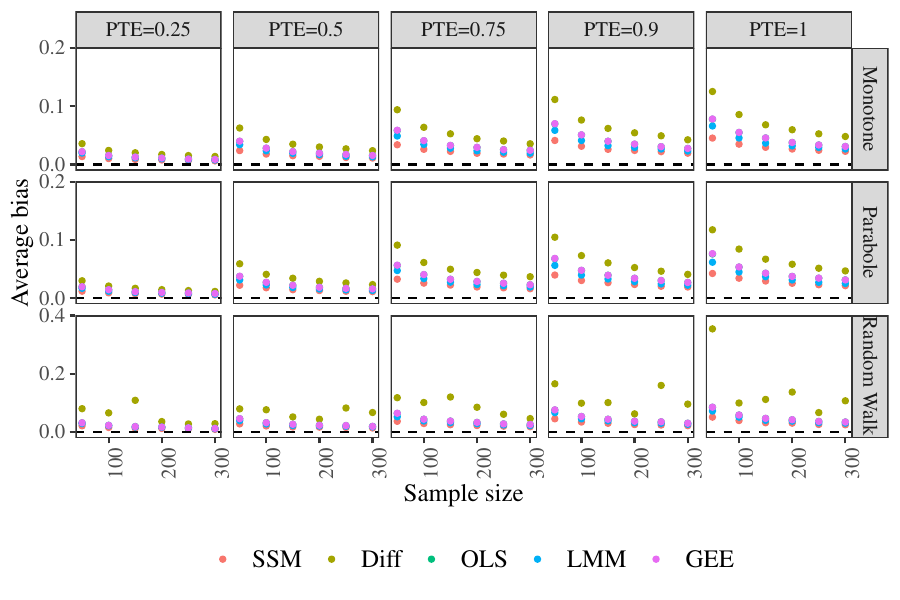}   
    \caption{Simulation results examining the average bias of the proposed SSM estimator (red), compared to estimation using GEE (green), LMM (blue), and OLS (purple), across 15 simulations which vary with respect to the treatment effect trajectory (monotone, parabole, or random walk) and the PTE (0.25, 0.5,0.75,0.9, or 1.0), and 6 sample sizes; a horizontal black dashed line is shown at a width of 0 for reference.}
    \label{fig:simul3_bias}
\end{figure}
\fi

With respect to a test for surrogate validity, we adopted the threshold $\PTE \ge0.75$, for a valid surrogate. Thus, testing a valid surrogate becomes equivalent to testing the following hypothesis:

\vspace*{-7mm}

$$\begin{aligned}
H_0: \PTE\le 0.75.\quad H_1: \PTE\ge 0.75,
\end{aligned}
$$

\noindent We carried out this test by checking whether the lower bound of the corresponding confidence interval for the PTE includes our threshold value of $0.75$. Specifically, we set our significance level to $5\%$ and compute the lower bound for the $90\%$ confidence interval for the PTE using each approach. Simulation results are shown in Figure  \ref{fig:simul3_power}. These results show that the Type I error for evaluating the surrogate is close to the nominal significance level, while providing non-trivial rejection rates when the true PTE is above the chosen threshold.

\if\figureend1
\begin{figure}
    \centering
    \includegraphics[scale=1]{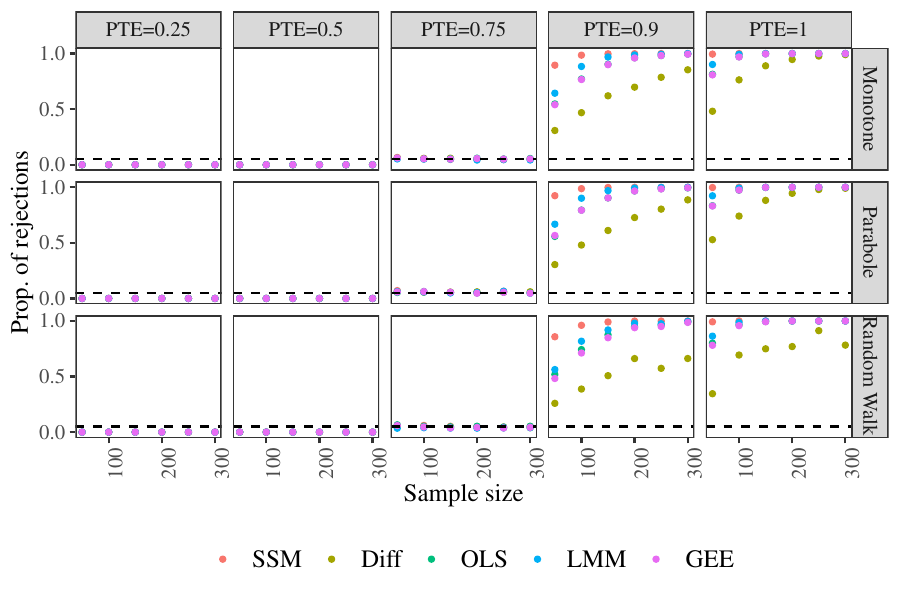}
    \caption{Simulation results examining the proportion of rejections of the null hypothesis that the PTE is $\leq 0.75$ using the proposed SSM estimator (red), compared to estimation using GEE (green), LMM (blue), and OLS (purple), across 15 simulations which vary with respect to the treatment effect trajectory (monotone, parabole, or random walk) and the PTE (0.25, 0.5,0.75,0.9, or 1.0), and 6 sample sizes; note that when the PTE $=$ 0.25 or 0.5 or 0.75, the null hypothesis is true and thus this calculated proportion is the empirical Type 1 error, with PTE$=0.75$ reflecting the boundary of the null hypothesis, and when the PTE = 0.9 or 1.0, the null hypothesis is false and thus, this proportion reflects the empirical power; a horizontal black dashed line is shown at a width of 0.05 for reference.}
    \label{fig:simul3_power}
\end{figure}
\fi

Overall, these results show that our proposed causal framework for surrogate evaluation works well in general across settings and that our proposed SSM estimation approach provides non-trivial improvements over reasonable comparator approaches.

\vspace*{-4mm}
\subsection{Results: Temporal Homogeneity}\label{subsec:test_eval}

Next, we investigated the performance of the proposed test for temporal homogeneity, described in Section \ref{subsec:hetero_test}.  As described in \ref{app:sims}, we examined five scenarios where Scenario 1 reflects the null setting in which the PTE did not vary over time, and Scenarios 2-5 were such that the PTE did vary over time. We compared our proposed MSD test to a Wald-type test, as this is a reasonable alternative to consider. Results are shown in Figure \ref{fig:simul7_power} which summarizes the empirical rejection probabilities for the MSD and Wald tests across all scenarios, sample sizes, and study lengths. In Scenario 1 (null setting), the proposed MSD test was well calibrated: its rejection rate remained close to $5\%$ for all sample sizes and follow-up durations. In contrast, the Wald test was approximately calibrated for short studies (up to 3 years) but became markedly anti-conservative as the follow-up length increased, with substantial size inflation for 5-year studies and severe inflation for 10-year studies. A plausible explanation is dimensionality: with measurements every 3 months, a 10-year study yields $T=40$ time points, such that the homogeneity hypothesis imposes equality across 40 time-specific parameters. In this regime, estimating the covariance structure needed for the Wald statistic can be unstable unless $N$ is very large relative to $T$. Because the Wald statistic aggregates deviations through a quadratic form, many noisy components can accumulate and inflate the test, whereas the proposed MSD statistic is driven by the largest standardized deviation and is therefore less sensitive to widespread moderate estimation error. Consistent with its asymptotic justification, the Wald test appears to require substantially larger samples to achieve reliable finite-sample size control when $T$ is large.

\if\figureend1
\begin{figure}
    \centering
    \includegraphics[scale=1]{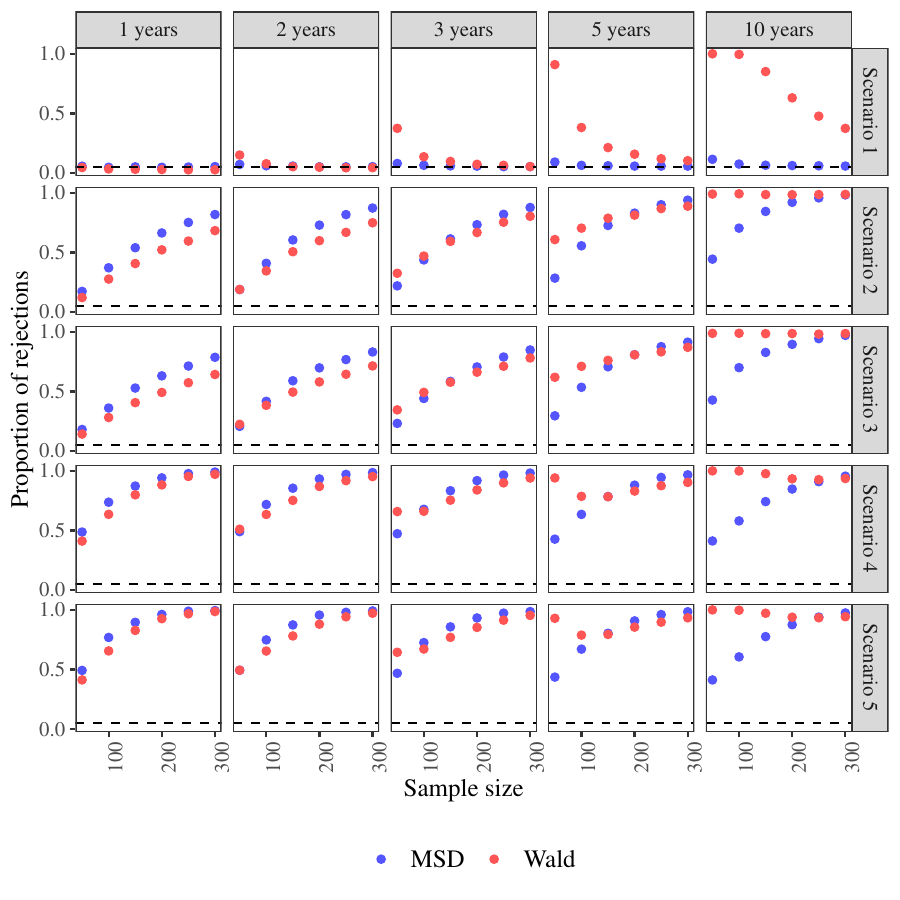}
    \caption{Simulation results examining the proportion of rejections of the null hypothesis that the PTE is constant over time i.e., temporal homogeneity, using the proposed MSD test compared to a Wald-based test at an $\alpha$-level of 0.05; note that the null hypothesis is true in Scenario 1 and is false in Scenarios 2, 3, 4, and 5; a horizontal black dashed line is shown at 0.05 for reference.}
    \label{fig:simul7_power}
\end{figure}
\fi

For Scenarios 2–5 (alternatives), when both procedures were reasonably calibrated (notably for shorter follow-up durations), power was similar, with MSD typically matching or slightly exceeding Wald. For longer studies, the Wald test’s higher rejection rates should be interpreted cautiously because they coincide with the size distortions observed under Scenario 1. Thus, based on these results, we recommend the MSD test for practical use due to its stable type I error control across all designs considered and its competitive power in settings where comparisons are meaningful.

\vspace*{-5mm}
\section{Application}\label{sec:app}

We applied our proposed method to investigate hemoglobin A1c (HbA1c) as a surrogate albumin excretion rate (AER) in the Diabetes Control and Complications Trial (DCCT) study. The DCCT study was multicenter, randomized, clinical study conducted about 1,441 patients with Type 1 diabetes. The study was designed to determine the effect of an intensive treatment protocol directed at maintaining blood glucose concentrations on progression of early vascular complications, compared to a control group that received usual care \citep{diabetes1993effect}. The surrogate marker of interest, HbA1c, is measured from blood samples and reflects blood glucose concentration. The outcome of interest, AER, measured in the urine, reflects kidney damage and is used to diagnose diabetic nephropathy, a microvascular complication of diabetes. In this study, HbA1c was routinely measured as part of the treatment monitoring protocol, thus, it was available frequently and non-invasively, whereas assessment of kidney damage progression via AER required additional testing via a urine sample. Trial participants were assessed every 3 months, thus reflecting our setting of interest with a longitudinal surrogate and a longitudinal outcome.

For our primary analysis, we fit the proposed model with up to 28 surrogate lags (7 years) and included two binary baseline covariates (sex and smoking status). While long lag structures are theoretically desirable, estimation at larger lags is limited in practice by attrition because fewer participants remain under follow-up long enough to contribute information. We therefore defined the practical maximum lag as the latest time point with at least 200 observations per treatment arm, which resulted in our 28-lag selection. 

Figure \ref{fig:cpte_28} shows the estimated CPTE trajectory for this 28-lag model along with the pointwise confidence intervals. At the final time point, the estimated CPTE was $0.874$ (90\% CI: $0.611$ to $1.276$). Although this point estimate suggests that change in HbA1c may explain a substantial proportion of the treatment effect on change in AER, because the lower bound of the confidence interval is not greater than 0.75, we would conclude that there is not sufficient evidence to conclude that it is a valid surrogate. Note that it is not surprising that Figure \ref{fig:cpte_28} shows confidence intervals with an upper bound greater than 1.0; this reflects finite-sample uncertainty rather than the estimand itself. Under assumptions \textbf{(S1)--(S3)}, the true PTE is bounded above by 1, but the estimator and its confidence interval are not necessarily constrained to lie in $[0,1]$. Our application of the proposed temporal homogeneity test resulted in a test statistic of $T=1.51$ and a p-value of $0.6458$, and thus we do not have evidence of heterogeneity in the PTE over time.  

\if\figureend1
\begin{figure}
\centering
    \begin{subfigure}[c]{0.49\textwidth}
        \centering
        \includegraphics[scale=1]{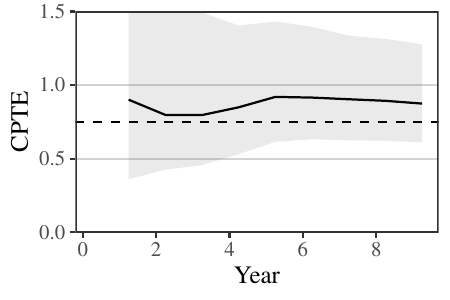}
        \caption{Estimation results using 28 lags }
        \label{fig:cpte_28}
    \end{subfigure}
    \begin{subfigure}[c]{0.49\textwidth}
        \centering
        \includegraphics[scale=1]{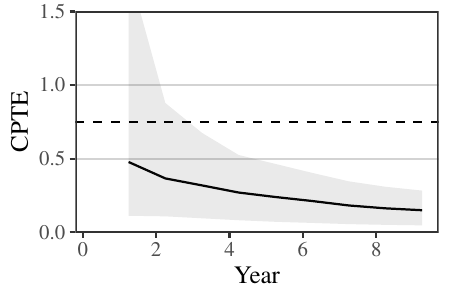}
        \caption{Estimation results using 0 lags}
        \label{fig:cpte_0}
    \end{subfigure}
     \caption{Diabetes clinical trial results examining change in  hemoglobin A1c as a surrogate for change in albumin excretion rate: the solid black line is the estimated cumulative proportion of treatment effect explained (CPTE) using our proposed methods over time using 28 lags (a) versus using 0 lags (b); the shaded region corresponds to estimated 90\% pointwise confidence intervals; the dashed black line is shown at a CPTE of 0.75 for reference.} \label{fig:cpte}
\end{figure}
\fi

To emphasize the importance of incorporating the surrogate \textit{trajectory}, rather than relying solely on the current surrogate value when assessing surrogate strength, Figure \ref{fig:cpte_0} presents the estimated CPTE trajectory using no lags (i.e., using only the current surrogate value). Comparing Figure \ref{fig:cpte_0} (no lags) with Figure \ref{fig:cpte_28} (28 lags) reveals substantial differences in both the estimated surrogate strength and its temporal heterogeneity. These findings underscore the importance of formally evaluating the longitudinal surrogate process, as enabled by our proposed framework. In \ref{app:diabetes}, we provide further model details and lag results, as well as a sensitivity analysis of the selection for the number of lags. 

\vspace*{-3mm}
\section{Discussion}\label{sec:conclusion}

Rigorous evaluation is essential before a surrogate can be considered as a replacement for a primary outcome in future trials. We have proposed a formal causal framework, based on a state-space approach, for settings in which both the surrogate and the primary outcome are observed longitudinally. Within this framework, we introduced local and cumulative versions of the PTE to characterize time-specific and aggregated measures of surrogacy. Our methodology leverages state-space models estimated via the Kalman filter, enabling flexible modeling of temporal dynamics while accommodating subject-level heterogeneity. We further developed a test for temporal homogeneity of surrogate strength and a nonparametric bootstrap procedure tailored to longitudinal trials. Simulation studies demonstrated strong finite-sample performance of the proposed methods, and application to a diabetes clinical trial illustrated their practical utility. The proposed methods are implemented in the R package \makeblind{\texttt{OnlineSurr}}, which is available on CRAN and at \makeblind{\url{https://github.com/silvaneojunior/OnlineSurr}}, which also contains a package tutorial and all reproducible code.

Though not unique to our approach, we require a number of both testable and untestable assumptions. The causal interpretation of the proposed PTE quantities and estimates depends on longitudinal analogs of standard surrogate identification conditions, as well as specified model conditions. In general, identification conditions cannot be empirically tested with available data. Regarding model specifications, the specific modeling strategy can accommodate complex effects and interactions including spline effects, treatment–surrogate interactions, regularized distributed-lag models, or rich lag structures, such as interactions among lagged surrogates. As is often the case in practice, certain complexities can be unstable in finite samples and may require careful control to preserve interpretability and stable inference. Beyond formal identification, it is also desirable to have some protection against the surrogate paradox, a situation in which the treatment effects on the surrogate and on the outcome are in opposite directions \citep{vanderweele2013surrogate,chen2007criteria}. In \ref{app:non-neg}, we discuss how standard sufficient conditions from the literature can be adapted to the longitudinal setting, while also proving that the proposed adaptations are, in fact, sufficient. Future work on developing sensitivity analyses that quantify robustness to violations of key identification assumptions in our setting as well as methods to inform principled lag selection and regularization would be useful. 

Finally, we focused on a randomized trial setting with a baseline, time-invariant binary treatment. Extending the framework to more complex treatment regimes is both important and nontrivial. In many applications, the surrogate, outcome, \textit{and treatment} may change over time, potentially in a stochastic or adaptive manner \citep{chakraborty2014dynamic,Diaz2023Mediation}. In such settings, the definition of the PTE would require careful reformulation to account for dynamic treatment regimes and time-dependent confounding. Moreover, the surrogate-specific objective of determining whether a surrogate can replace the primary outcome in capturing the treatment effect becomes substantially more intricate. Although this complexity does not arise in our diabetes application, it may be relevant in clinical trials where treatment is deliberately modified by design. Developing extensions of the present framework to support surrogate evaluation in a stochastic treatment  setting represents a valuable direction for future research. 

\vspace*{-5mm}
\section*{Acknowledgments}
\makeblind{This work was supported by NIDDK grant R01DK118354 (PI:Parast).The Diabetes Control and Complications Trial (DCCT) and its follow-up the Epidemiology of Diabetes Interventions and Complications (EDIC) study were conducted by the DCCT/EDIC Research Group and supported by National Institutes of Health (NIH) grants and contracts and by the General Clinical Research Center Program (GCRC), the National Center for Research Resources (NCRR). The resources from the DCCT/EDIC study were supplied by NIDDK Central Repository (NIDDK-CR). This manuscript was not prepared under the auspices of the DCCT/EDIC study and does not represent analyses or conclusions of the DCCT/EDIC study group, NIDDK-CR, or NIH.
}

\section*{Data Availability Statement}

The DCCT/EDIC data are available by request from the NIDDK Central Repository at \url{https://repository.niddk.nih.gov/}; the study resource DOI is \url{https://doi.org/10.58020/mn8k-ms49}. Reproducible code is available at \makeblind[\url{https://anonymous.4open.science/r/reproductible-codes-02B5}]{\url{https://github.com/silvaneojunior/OnlineSurr}}.

\vspace*{-5mm}
\section*{Disclosure}
The authors report there are no
competing interests to declare.
\vspace*{-5mm}
\clearpage
\if\figureend0
\begin{figure}
    \centering
    \includegraphics[scale=1]{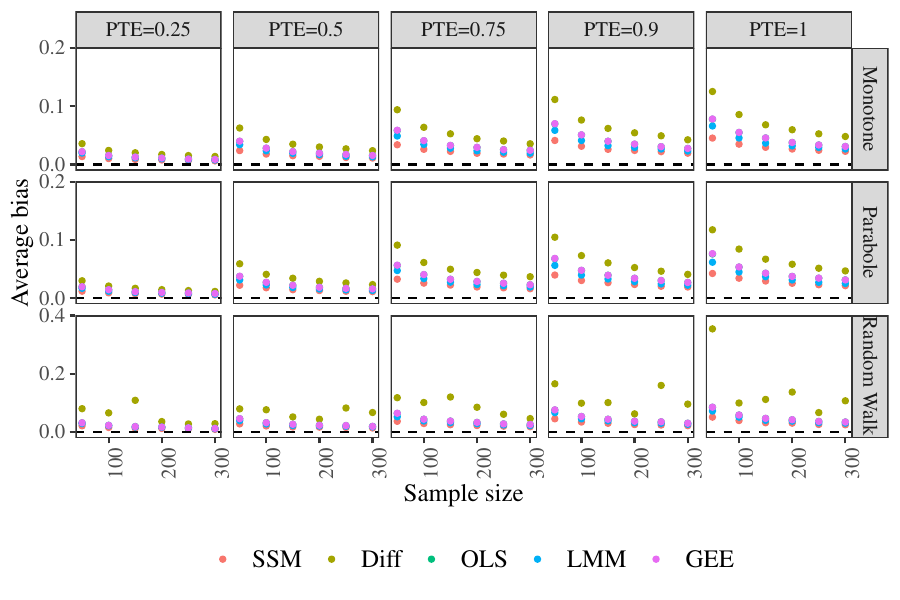}   
    \caption{Simulation results examining the average bias of the proposed SSM estimator (red), compared to estimation using GEE (green), LMM (blue), and OLS (purple), across 15 simulations which vary with respect to the treatment effect trajectory (monotone, parabole, or random walk) and the PTE (0.25, 0.5,0.75,0.9, or 1.0), and 5 sample sizes; a horizontal black dashed line is shown at 0 for reference.}
    \label{fig:simul3_bias}
\end{figure}

\begin{figure}
    \centering
    \includegraphics[scale=1]{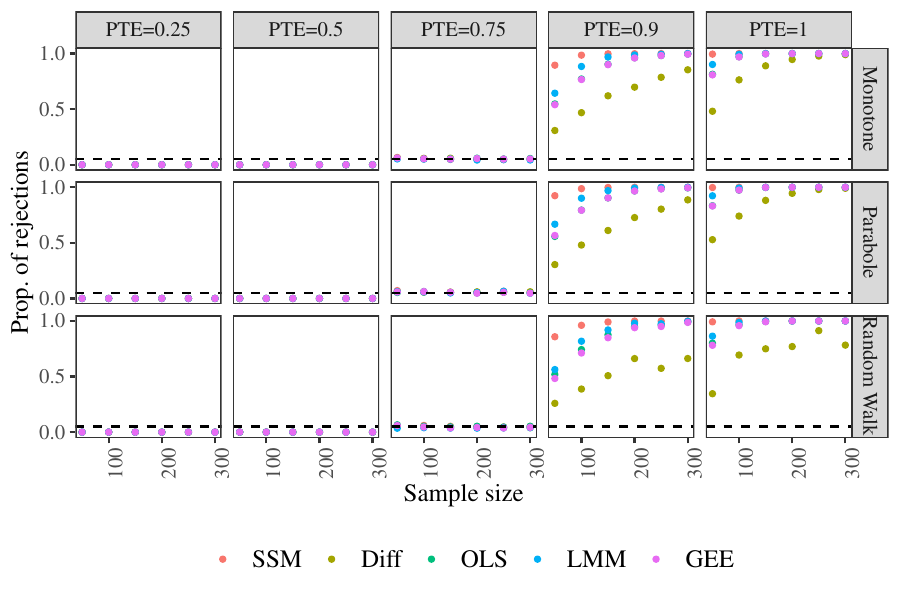}
    \caption{Simulation results examining the proportion of rejections of the null hypothesis that the PTE is $\leq 0.75$ using the proposed SSM estimator (red), compared to estimation using GEE (green), LMM (blue), and OLS (purple), across 15 simulations which vary with respect to the treatment effect trajectory (monotone, parabole, or random walk) and the PTE (0.25, 0.5,0.75,0.9, or 1.0), and 5 sample sizes; note that when the PTE $=$ 0.25 or 0.5 or 0.75, the null hypothesis is true and thus this calculated proportion is the empirical Type 1 error, with PTE$=0.75$ reflecting the boundary of the null hypothesis, and when the PTE = 0.9 or 1.0, the null hypothesis is false and thus, this proportion reflects the empirical power; a horizontal black dashed line is shown at 0.05 for reference.}
    \label{fig:simul3_power}
\end{figure}

\begin{figure}
    \centering
    \includegraphics[scale=1]{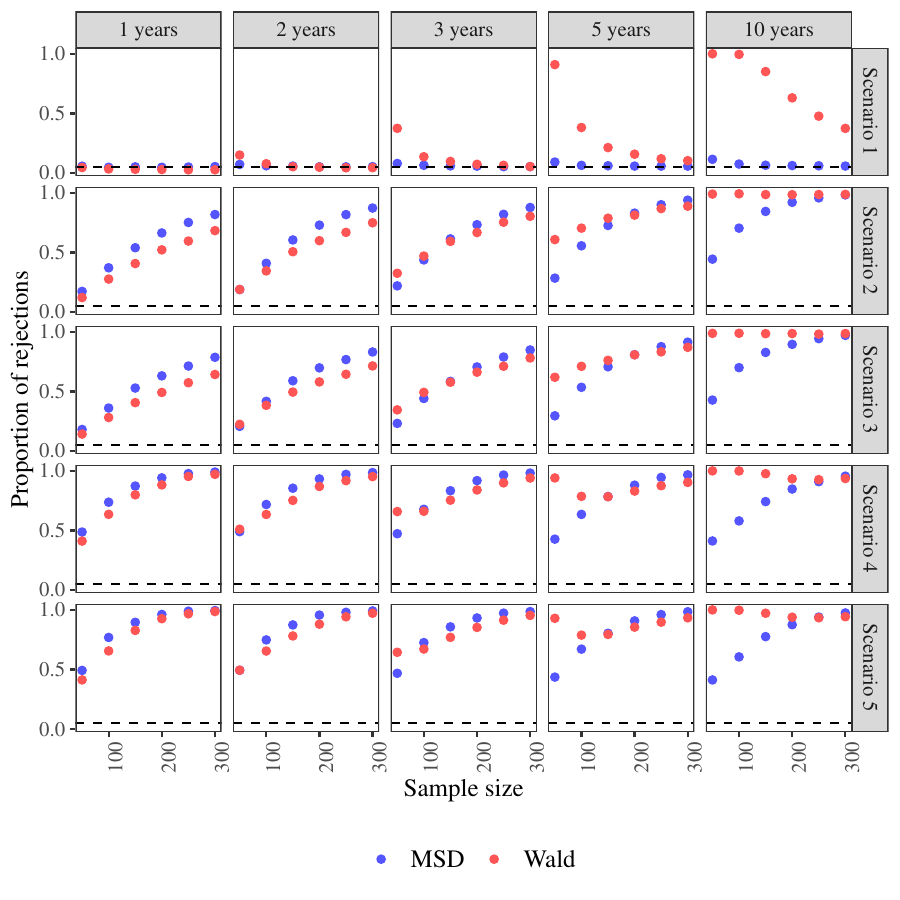}
    \caption{Simulation results examining the proportion of rejections of the null hypothesis that the PTE is constant over time i.e., temporal homogeneity, using the proposed MSD test compared to a Wald-based test at an $\alpha$-level of 0.05; note that the null hypothesis is true in Scenario 1 and is false in Scenarios 2, 3, 4, and 5; a horizontal black dashed line is shown at 0.05 for reference.}
    \label{fig:simul7_power}
\end{figure}

\clearpage
\begin{figure}
\centering
    \begin{subfigure}[c]{0.49\textwidth}
        \centering
        \includegraphics[scale=1]{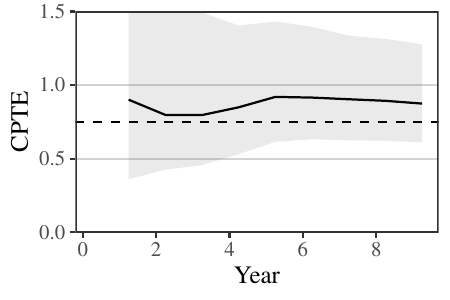}
        \caption{Estimation results using 28 lags }
        \label{fig:cpte_28}
    \end{subfigure}
    \begin{subfigure}[c]{0.49\textwidth}
        \centering
        \includegraphics[scale=1]{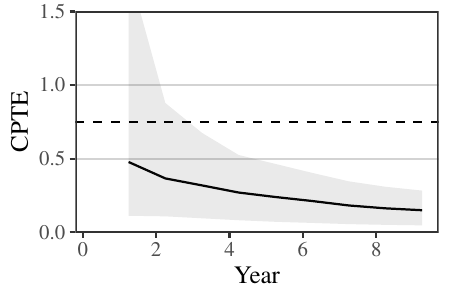}
        \caption{Estimation results using 0 lags}
        \label{fig:cpte_0}
    \end{subfigure}
     \caption{Diabetes clinical trial results examining change in  hemoglobin A1c as a surrogate for change in albumin excretion rate: the solid black line is the estimated cumulative proportion of treatment effect explained (CPTE) using our proposed methods over time using 28 lags (a) versus using 0 lags (b); the shaded region corresponds to estimated 90\% pointwise confidence intervals; the dashed black line is shown at a CPTE of 0.75 for reference.} \label{fig:cpte}
\end{figure}
\fi

\clearpage
\appendix

\setcounter{table}{0}
\setcounter{table}{0}
\renewcommand{\thetable}{A\arabic{table}}
\renewcommand{\thesection}{Appendix \Alph{section}}

\setcounter{figure}{0}
\renewcommand{\thefigure}{A\arabic{figure}}
\renewcommand{\theequation}{A.\arabic{equation}}

\section{} 

\subsection{Additional Conditions \label{app:non-neg}}

{
While the identifiability condition \textbf{(D4)} is sufficient for estimating $\delta_{2,t}$, by itself this condition does not guarantee that the PTE can be interpreted as a proportion in the traditional sense. Specifically, if $\delta_{2,t}$ is negative at some time $t$, the PTE may not lie in $[0,1]$, thereby compromising its interpretation as a ``proportion.'' More importantly, this could lead to the surrogate paradox, a situation in which the treatment effects on the surrogate and on the outcome are in opposite directions \citep{vanderweele2013surrogate,chen2007criteria}. Thus, one may wish to impose conditions that guarantee that the surrogate paradox does not occur. This can be achieved through the following sufficient, though not necessary, additional conditions:

\begin{enumerate}
    \item[\textbf{(S1)}]  \textbf{Non-negative residual treatment effect}:
    $    \mathbb{E}\left[Y^{(1)}_t\middle|\bS^{(1)}_t=\boldsymbol{s}\right]\ge \mathbb{E}\left[Y^{(0)}_t\middle|\bS^{(0)}_t=\boldsymbol{s}\right],
    $
    for all times $t$ and all trajectories $\boldsymbol{s}$.
    \item[\textbf{(S2)}]  \textbf{Non-negative  surrogate effect}:
    $\mathbb{E}[Y^{(g)}_t|S^{(g)}_{t}=s_t,\cdots,S^{(g)}_{0}=s_{0}]$ is a monotone non-decreasing function of $s_i, i=0,\cdots, t,$ for both treatment arms $g$.
    \item[\textbf{(S3)}] \textbf{Forward stochastic dominance}:
    $\mathbb{P}(S^{(1)}_t>s_t|\bS^{(1)}_{t-1}=\boldsymbol{s})\ge \mathbb{P}(S^{(0)}_t>s_t|\bS^{(0)}_{t-1}=\boldsymbol{s}),$ for all trajectories $\boldsymbol{s}$ and for all times $t$.
\end{enumerate}

These conditions are not unique to our work, as they are analogous to those commonly used in surrogate evaluation methods \citep{wang2002measure,vanderweele2013surrogate}. In this respect, our contribution lies in making the necessary adaptations for the longitudinal setting, while also proving that the proposed version of the assumptions guarantees that $\delta_{2,t}$ is non-negative for all $t$. Notably, in the longitudinal setting Assumptions \textbf{(S1)--(S3)} require conditioning on the entire past trajectory. 

Note that \textbf{(S2)} allows for $\mathbb{E}[Y^{(G)}_t|S^{(G)}_{t}=s_t,\cdots,S^{(G)}_{1}=s_{1}]$ to be constant in $s_i$; that is, this does not require that all previous surrogates have an effect on the current measurement of the outcome, only that they do not have opposite directions. 

Lastly, Assumption \textbf{(S3)} requires that conditional on the surrogate history, $S\supone_t$ stochastically dominates $S\supzero_t$, which implies that the effect of the treatment on the surrogate has a non-negative direction throughout time, though notably the effect can be zero.

It is worth noting that these additional conditions are not, by any means, trivial, nor should they be taken for granted without careful consideration. Previous work including \cite{elliott2015surrogacy,hsiao2025avoiding,hsiao2025resilience} has offered empirical methods to test these assumptions and evaluate sensitivity to their violations in the non-longitudinal setting. Extensions to the longitudinal setting are non-trivial and constitute a natural direction for expanding the current work.

}

We now show that Conditions \textbf{(S1)--(S3)} imply that $\delta_{2,t}\geq 0$ for all $t$. Suppose that \textbf{(S1)--(S3)} hold. In addition, suppose that (we will prove this afterwards)
\begin{equation}
\int \mathbb{E}[Y_t|G=0,\bS_t=\boldsymbol{s}] dF_{\bS\supone_t}(\boldsymbol{s})
\geq
\int \mathbb{E}[Y_t|G=0,\bS_t=\boldsymbol{s}] dF_{\bS\supzero_t}(\boldsymbol{s}).
\label{eq:induc_step1}
\end{equation}

\noindent Then, it follows that 
\begin{eqnarray}
\delta_{2,t} &=& \mathbb{E}[Y_{t}|G=1]-\mathbb{E}[Y_{t}|G=0] \nonumber\\
&=& \int \mathbb{E}[Y_{t}|G=1,\bS_t=\boldsymbol{s}] dF_{\bS\supone_t}(\boldsymbol{s})
-\int \mathbb{E}[Y_{i,t}|G_i=0,\bS_t=\boldsymbol{s}] dF_{\bS\supzero_t}(\boldsymbol{s}) \nonumber\\
&\geq& \int \mathbb{E}[Y_{t}|G=0,\bS_t=\boldsymbol{s}] dF_{\bS\supone_t}(\boldsymbol{s})
-\int \mathbb{E}[Y_{i,t}|G_i=0,\bS_t=\boldsymbol{s}] dF_{\bS\supzero_t}(\boldsymbol{s}), \label{uses2}\\
&\geq& 0, \label{uses3}
\end{eqnarray}

\noindent where $\boldsymbol{s}=(s_t,\ldots,s_1)$, and (\ref{uses2}) follows from \textbf{(S1)}, and (\ref{uses3}) follows from (\ref{eq:induc_step1}). Thus, our aim is to prove (\ref{eq:induc_step1}). We prove this claim by induction on the sequence of nested integrals, additionally relying on Conditions \textbf{(S2)} and \textbf{(S3)}. First, we rewrite the left-hand side as
$$
\begin{aligned}
\int \mathbb{E}[Y_t|G=0,\bS_t=\boldsymbol{s}] dF_{\bS\supone_t}(\boldsymbol{s})
&=
\int \cdots \int \mathbb{E}[Y_t|G=0,S_t=s_t,\ldots,S_1=s_1] \\
&\qquad\qquad dF_{S\supone_t|\bS\supone_{t-1}}(s_t)\cdots dF_{S\supone_1}(s_1).
\end{aligned}
$$

\noindent We now show, moving from the innermost integral outward, that each replacement of $F_{S\supone_j|\bS\supone_{j-1}}$ by $F_{S\supzero_j|\bS\supzero_{j-1}}$ preserves the inequality.

For the innermost integral, \textbf{(S2)} implies that
$\mathbb{E}[Y_t|G=0,S_t=s_t,\ldots,S_1=s_1]$ is monotone in $s_t$. Hence, by \textbf{(S3)},
$$
\begin{aligned}
\int& \mathbb{E}[Y_t|G=0,S_t=s_t,\ldots,S_1=s_1]
dF_{S\supone_t|\bS\supone_{t-1}}(s_t)\\
&\geq
\int \mathbb{E}[Y_t|G=0,S_t=s_t,\ldots,S_1=s_1]
dF_{S\supzero_t|\bS\supzero_{t-1}}(s_t).
\end{aligned}
$$

\noindent This establishes the base step. Next, suppose that for some $k\geq 0$,
$$
\begin{aligned}
\int &\cdots \int \mathbb{E}[Y_t|G=0,S_t=s_t,\ldots,S_1=s_1]
dF_{S\supone_t|\bS\supone_{t-1}}(s_t)\cdots dF_{S\supone_{t-k}|\bS\supone_{t-k-1}}(s_{t-k})
\\
&\geq
\int \cdots \int \mathbb{E}[Y_t|G=0,S_t=s_t,\ldots,S_1=s_1]
dF_{S\supzero_t|\bS\supzero_{t-1}}(s_t)\cdots dF_{S\supzero_{t-k}|\bS\supzero_{t-k-1}}(s_{t-k}).
\end{aligned}
$$

\noindent Integrating both sides with respect to $dF_{S\supone_{t-k-1}|\bS\supone_{t-k-2}}(s_{t-k-1})$ yields
$$
\begin{aligned}
\int &\int \cdots \int \mathbb{E}[Y_t|G=0,S_t=s_t,\ldots,S_1=s_1]dF_{S\supone_t|\bS\supone_{t-1}}(s_t)\cdots dF_{S\supone_{t-k}|\bS\supone_{t-k-1}}(s_{t-k})
 \\&dF_{S\supone_{t-k-1}|\bS\supone_{t-k-2}}(s_{t-k-1})
\\
&\geq
\int \int \cdots \int \mathbb{E}[Y_t|G=0,S_t=s_t,\ldots,S_1=s_1]
dF_{S\supzero_t|\bS\supzero_{t-1}}(s_t)\cdots dF_{S\supzero_{t-k}|\bS\supzero_{t-k-1}}(s_{t-k})
\\ & dF_{S\supone_{t-k-1}|\bS\supone_{t-k-2}}(s_{t-k-1}).
\end{aligned}
$$

\noindent By \textbf{(S2)}, the inner expression
$$
\int \cdots \int \mathbb{E}[Y_t|G=0,S_t=s_t,\ldots,S_1=s_1]
dF_{S\supzero_t|\bS\supzero_{t-1}}(s_t)\cdots dF_{S\supzero_{t-k}|\bS\supzero_{t-k-1}}(s_{t-k})
$$
is monotone in $s_{t-k-1}$. Therefore, applying \textbf{(S3)} once more gives
$$
\begin{aligned}
\int &\int \cdots \int \mathbb{E}[Y_t|G=0,S_t=s_t,\ldots,S_1=s_1]
dF_{S\supone_t|\bS\supone_{t-1}}(s_t)\cdots dF_{S\supone_{t-k}|\bS\supone_{t-k-1}}(s_{t-k})\\
 &dF_{S\supone_{t-k-1}|\bS\supone_{t-k-2}}(s_{t-k-1})
\\
&\geq
\int \int \cdots \int \mathbb{E}[Y_t|G=0,S_t=s_t,\ldots,S_1=s_1]
dF_{S\supzero_t|\bS\supzero_{t-1}}(s_t)\cdots dF_{S\supzero_{t-k}|\bS\supzero_{t-k-1}}(s_{t-k})\\
 &dF_{S\supzero_{t-k-1}|\bS\supzero_{t-k-2}}(s_{t-k-1}).
\end{aligned}
$$

\noindent This completes the induction, proving (\ref{eq:induc_step1}), and thus, 
 $\delta_{2,t}\geq 0$ for all $t$.

\subsection{Model conditions: details \label{app:conditions} }

\noindent Condition \textbf{(C1)} is associated with the design of the study, and should generally hold for randomized trials.
If the first part of condition \textbf{(C2)} does not hold, then the PTE is simply not properly defined. If the second part of   condition \textbf{(C2)} does not hold, then we would have an infinite variance estimator that, while even if unbiased, does not converge as the sample size increases.

Condition \textbf{(C3)} is a standard weak-dependence condition. It is automatically satisfied when subject-level trajectories are independent, which is the usual working assumption in individually randomized trials. However, \textbf{(C3)} does not require independence; it only requires that such cross-subject dependence is weak enough for the empirical estimating equations to converge. In practice, this condition can often be justified by the trial design. For example, it is reasonable in individually randomized trials with independent recruitment and no persistent common recruitment/trial disruptions. If relevant cross-subject dependence is expected, such as clustering by site or clinic, it can be incorporated into the working covariance structure or addressed by defining the asymptotics at the cluster level rather than the subject level. Misspecifying this dependence structure may reduce efficiency, but it does not affect consistency as long as the average cross-subject dependence vanishes asymptotically, as required in \textbf{(C3)}.

Note that, once \textbf{(C1)}--\textbf{(C3)} hold, condition \textbf{(C4)} is primarily a stability condition on the chosen working GLS/SSM specification. In the proposed SSM, a sufficient way to ensure the stability component of \textbf{(C4)} is to define the working precision matrix $A_m$ through a discount strategy with discount factors bounded away from both 0 and 1, that is:

$$
0<\epsilon_1 \leq \delta_t \leq \epsilon_2<1,
$$
together with a finite and nondegenerate initial covariance matrix. Importantly, this condition concerns only the working covariance matrices used by the Kalman recursions, not the true covariance structure of the data. In particular, under the discount strategy, the point estimator depends on the relative scale of the state and observation variances. Thus, the working observation variance can be fixed at an arbitrary positive value, such as the identity matrix. The initial covariance matrix is then a modeling choice, playing the role of a regularization parameter, and should be chosen to be finite and nondegenerate.

Misspecification of the working observation variance may affect efficiency, particularly if the true variance is time-varying, and may also affect model-based uncertainty quantification. However, it does not affect consistency of the fixed-effect estimator, provided that the induced working precision matrices remain stable. In our procedure, uncertainty quantification is obtained through the nonparametric bootstrap, rather than through the model-based Kalman variance. Thus, misspecification of the working observation variance is not a concern for the proposed inferential procedure under the stated target and stability conditions.

Lastly, time-varying variances can be incorporated into the working model to improve efficiency. In that case, the method used to model heteroskedasticity must preserve \textbf{(C4)}. For example, \cite{West-DLM} provide discount-based approaches for modeling time-varying variances, and \cite{Alves-kparametric} propose a related approach; both are compatible with the required regularity conditions when the resulting working covariance matrices remain bounded and nondegenerate.

Thus, under finite second moments, bounded discount factors, a finite and nondegenerate initial covariance matrix, and a fixed positive working observation variance, the covariance matrices generated by the Kalman recursions remain bounded and nondegenerate for fixed $T$. Combined with the treatment-arm proportion condition, which guarantees that the fixed-effect design matrix is well behaved in the simplified no-covariate setting, this implies \textbf{(C4)}. For the more general setting, the covariates must satisfy standard regularity conditions, such as being linear independent.

{
Condition \textbf{(C5)} is related to the final form of the estimator presented in Section \ref{subsec:PTE}, which employs a Monte Carlo approximation. By \textbf{(C2)}, we have that:

$$
 \text{Var}\left[Y_t^{(1)}\right] =  \mathbb{E}\left[\text{Var}\left[Y_t^{(1)}\middle| \bS_t\right]\right] +  \text{Var}\left[\mathbb{E}\left[Y_t^{(1)}\middle| \bS_t\right]\right]<\infty
$$

\noindent Thus, since $\text{Var}\left[\mathbb{E}\left[Y_t^{(1)}\middle| \bS_t\right]\right]<\infty$, we have that:

$$
 \mathbb{E}\left[\mathbb{E}\left[Y_t^{(1)}\middle| \bS_t\right]^2\right] = \int \mathbb{E}\left[Y_t^{(1)}\middle|\bS^{(1)}_t=\boldsymbol{s}\right]^2dF_{\bS\supone}(\boldsymbol{s})<\infty
$$

\noindent This is quite similar to \textbf{(C5)}, except that the integral under the probability measure here is $F_{\bS\supone}$ instead of $F_{\bS\supzero}$. If conditions \textbf{(S2)} and \textbf{(S3)} hold, which are the conditions discussed in \ref{app:non-neg} to protect against the surrogate paradox, then we have that:

$$
  \int \mathbb{E}\left[Y_t^{(1)}\middle|\bS^{(1)}_t=\boldsymbol{s}\right]^2dF_{\bS\supzero}(\boldsymbol{s}) \le \int \mathbb{E}\left[Y_t^{(1)}\middle|\bS^{(1)}_t=\boldsymbol{s}\right]^2dF_{\bS\supone}(\boldsymbol{s})<\infty.
$$

\noindent Thus, under \textbf{(S2)--(S3)}, \textbf{(C2)} implies \textbf{(C5)}.
}

{
\subsection{Pseudo-true limits \label{app:pseudo} }
Here, we detail the pseudo-true limits defined by the working mean structure, $\widetilde{\Delta}_{R}(t)$ and $\widetilde{\Delta}(t)$, assuming that conditions \textbf{(C1)--(C5)} hold. From this, we can derive condition \textbf{(C6)} and introduce a bias correction that guarantees it holds even under model misspecification. We consider the case where all previous surrogate effects are included in the model. The case of misspecification by the omission of lingering effects is related to the causal interpretation of the estimators, which is discussed in detail in Section \ref{subsec:lingering} of the main text.

In this section, we first show that, in the subject-level homoskedastic model, as long as the marginal model contains an intercept, randomization guarantees consistency of the treatment-effect estimator, even when the working mean structure is misspecified. Next, we consider the conditional model where, because the surrogate is not randomized, additional work is needed to establish consistency. When the working conditional mean is correctly specified, we show below that the estimator is consistent. We also show that, when it is misspecified, consistency can still be recovered through an appropriate bias-correction procedure. These particular results are not new and thus, we refer the reader to \cite{Wager2016} for the first result and to \cite{SHIMODAIRA2000227} and \cite{Steingrimsson2022} for the second. However, we do develop the specific bias-correction calculations which are needed for our specific setting.

Finally, we discuss how the subject-level heteroskedastic model differs from the homoskedastic case and show that the natural extension of the bias correction to this setting is equivalent to using the homoskedastic model's estimator. Therefore, unless the working mean model is correctly specified, the subject-level homoskedastic model is preferred, since it provides a consistent estimator of the PTE even under working mean misspecification and even when the true data-generating process exhibits subject-level heteroskedasticity. It is also important to note that time-varying heteroskedasticity is not inherently problematic; rather, subject-level heteroskedasticity is the specific concern, because it can act as a confounding factor for the treatment effect.

To characterize the pseudo-true limits in Theorem 1, we consider the subject outcome trajectory $\bY_{i} = (Y_{i,1},\cdots,Y_{i,t})$,  surrogate trajectory $\bS_{i} = (S_{i,1},\cdots,S_{i,t})$ and covariate vector $\bX_i$, with the sampling units being $(\bY_i,\bS_i,\bX_i)$. Since $T$ is fixed, the fixed-effects block of the SSM/Kalman estimator has the same large-sample target as the corresponding GLS normal equations. Let $\boldsymbol{m}(\bS_i,\bX_{i},G_i;\boldsymbol{\eta})=(m_1(\bS_i,\bX_{i},G_i;\boldsymbol{\eta}),\cdots,m_T(\bS_i,\bX_{i},G_i;\boldsymbol{\eta}))^{\top}$ denote a general working mean model, where $\boldsymbol{\eta}$ is a vector of unknown parameters. Also, for an analytic function $\boldsymbol{f}:\mathbb{R}^{n} \rightarrow R^{m}$, we denote its Jacobian as:
$$
J_{\boldsymbol{f}}(\boldsymbol{x_0})=\frac{\partial}{\partial \boldsymbol{x}^{\top}}\boldsymbol{f}(\boldsymbol{x_0}) \in \mathbb{R}^{m \times n},
$$

\noindent and we denote $\bepsilon_{i}(\boldsymbol{\eta})=\bY_{i}-\boldsymbol{m}(\bS_i,\bX_{i},G_i;\boldsymbol{\eta})$. Let $A$ be a $T\times T$ matrix representing the weights implied by the temporal evolution used on the Kalman filter, which we consider as known for this section. Then, the probability limit of $\boldsymbol{\eta}$, denoted $\boldsymbol{\eta}^*$, is characterized by:
\begin{equation}
\mathbb{E}\left[J_{\boldsymbol{m}}(\bS_i,\bX_{i},G_i;\boldsymbol{\eta^{*}})^{\top}A\bepsilon_i(\boldsymbol{\eta^{*}})
\right]=\boldsymbol{0}, \label{problimit}
\end{equation}

\noindent where $\bepsilon_i^{*}= \bepsilon_i(\boldsymbol{\eta}^{*})$. One solution for \ref{problimit} is obtained by setting 
$\boldsymbol{m}(\bS_i,\bX_{i},G_i;\boldsymbol{\eta^{*}}) = \mathbb{E}[\bY_i | \bS_i,\bX_{i},G_i]$. If the functional form of $\boldsymbol{m}$ is such that no $\boldsymbol{\eta}$ satisfies this equality, we say the model is misspecified. In this case, $\boldsymbol{\eta}^*$ is still well defined and we refer to it as the pseudo-true limit for $\boldsymbol{\eta}$. Below, we aim to derive the general form of $\boldsymbol{\eta}^*$.


\subsubsection*{Marginal model}

Consider the following covariate-adjusted working marginal model:
$$
\boldsymbol{m}(\bX_{i},G_i;\boldsymbol{\mu},\boldsymbol{\beta},\boldsymbol{\delta})=\boldsymbol{\mu}+\boldsymbol{g}(\bX_{i};\boldsymbol{\beta})+\boldsymbol{\delta} G_i,
$$
where $\boldsymbol{\mu}=(\mu_1,\cdots,\mu_T)$ represents the local level, $\boldsymbol{\delta}=(\delta_1,\cdots,\delta_T)$ represents the treatment effect and $\boldsymbol{g}(\bX_{i};\boldsymbol{\beta})=(g_1(\bX_{i};\boldsymbol{\beta}),\cdots,g_T(\bX_{i};\boldsymbol{\beta}))^{\top}$ represents the covariate effect. It is easy to see that, in this setting, $\boldsymbol{\eta} = (\boldsymbol{\mu}^{\top},\boldsymbol{\beta}^{\top},\boldsymbol{\delta}^{\top})^{\top}$. When convenient, we also write $m_t(\bX_{i},G_i;\boldsymbol{\mu},\boldsymbol{\beta},\boldsymbol{\delta})=\mu_t+g_t(\bX_{i};\boldsymbol{\beta})+\delta_t G_i, $to represent the working mean at time $t$. Lastly, it is important to note that $\bX_{i}$ does \textbf{not} include the main treatment indicator $G_i$. 

Then, by (\ref{problimit}) the pseudo-true values $\boldsymbol{\mu}^*,\boldsymbol{\beta}^*,\boldsymbol{\delta}^*$ solve:
\begin{eqnarray}
\mathbb{E}\left[A\left(\bY_{i}-\boldsymbol{\mu}^{*}-\boldsymbol{g}(\bX_{i};\boldsymbol{\beta}^*)-\boldsymbol{\delta}^* G_i\right)\right]&=&\boldsymbol{0}  \label{eq1}\\
\mathbb{E}\left[J_g(\bX_{i};\boldsymbol{\beta}^*)^{\top}A\left(\bY_{i}-\boldsymbol{\mu}^{*}-\boldsymbol{g}(\bX_{i};\boldsymbol{\beta}^*)-\boldsymbol{\delta}^* G_i\right)\right]&=&\boldsymbol{0}  \label{eq2}\\
\quad\mathbb{E}\left[G_iA\left(\bY_{i}-\boldsymbol{\mu}^{*}-\boldsymbol{g}(\bX_{i};\boldsymbol{\beta}^*)-\boldsymbol{\delta}^* G_i\right)\right]&=&\boldsymbol{0}  \label{eq3}.
\end{eqnarray}

\noindent Equation (\ref{eq1}) implies:
\begin{equation}
A \boldsymbol{\mu}^{*}=\mathbb{E}\left[A\bY_{i}\right]-\mathbb{E}\left[A\boldsymbol{g}(\bX_{i};\boldsymbol{\beta}^*)\right]-A\boldsymbol{\delta}^*\rho \label{eq1implies}
\end{equation}
\noindent  where $\rho = P(G_i=1)$ and by \textbf{(D2)}, $0<\rho<1$. Equation (\ref{eq3}) implies: 
$$
A\boldsymbol{\delta}^*=A\mathbb{E}\left[\bY_{i}-\boldsymbol{\mu}^*-\boldsymbol{g}(\bX_{i};\boldsymbol{\beta}^*)|G_i=1\right]=A\left(\mathbb{E}[\bY_{i}|G_i=1]-\boldsymbol{\mu}^*-\mathbb{E}[\boldsymbol{g}(\bX_{i};\boldsymbol{\beta}^*)|G_i=1]\right).
$$

\noindent Combining this result with (\ref{eq1implies}):
$$
A\boldsymbol{\delta}^*=\mathbb{E}[A\bY_{i}|G_i=1]-\mathbb{E}\left[A\bY_{i}\right]+\mathbb{E}\left[A\boldsymbol{g}(\bX_{i};\boldsymbol{\beta}^*)\right]+A\boldsymbol{\delta}^*\rho-\mathbb{E}[A\boldsymbol{g}(\bX_{i};\boldsymbol{\beta}^*)|G_i=1].
$$

\noindent Rearranging terms, it follows that:

$$
A\boldsymbol{\delta}^*=A\frac{-\mathbb{E}\left[\bY_{i}\right]+\mathbb{E}\left[\boldsymbol{g}(\bX_{i};\boldsymbol{\beta}^*)\right]+\mathbb{E}[\bY_{i}|G_i=1]-\mathbb{E}[\boldsymbol{g}(\bX_{i};\boldsymbol{\beta}^*)|G_i=1]}{1-\rho}.
$$

\noindent Noting that by randomization, $\mathbb{E}[\boldsymbol{g}(\bX_{i};\boldsymbol{\beta}^*)|G_i=1]=\mathbb{E}[\boldsymbol{g}(\bX_{i};\boldsymbol{\beta}^*)]$, as $\bX_{i}$ has the same distribution in the control and treatment groups, and that $\mathbb{E}[\bY_{i}] = (1-\rho)\mathbb{E}[\bY_{i}|G_i=0] + \rho\mathbb{E}[\bY_{i}|G_i=1]$, it follows that: 

$$
\begin{aligned}
A\boldsymbol{\delta}^*&=A\frac{\mathbb{E}[\bY_{i}|G_i=1]-(1-\rho)\mathbb{E}[\bY_{i}|G_i=0] - \rho\mathbb{E}[\bY_{i}|G_i=1]}{1-\rho}\\
&=A\left(\mathbb{E}[\bY_{i}|G_i=1]-\mathbb{E}[\bY_{i}|G_i=0]\right).
\end{aligned}
$$

\noindent Lastly, by condition \textbf{(C4)}, $A$ is non-singular, thus:

$$
\begin{aligned}
\boldsymbol{\delta}^*&=\mathbb{E}[\bY_{i}|G_i=1]-\mathbb{E}[\bY_{i}|G_i=0].
\end{aligned}
$$

\noindent That is, $\widehat{\Delta}(t)$ is a consistent estimator of $\Delta(t)$ without further assumptions, similar to the results presented in \cite{Wager2016}.

\subsubsection*{Conditional model}

We now study the bias induced by potential misspecification of the conditional model. We omit baseline covariates from the notation below for simplicity. Let $\bS_{i}$ denote the surrogate trajectory included in the conditional model, and consider the working conditional mean as:
$$
\boldsymbol{m}(\bX_{i},\bS_{i},G_i;\boldsymbol{\mu},\boldsymbol{\beta},\boldsymbol{\delta})=\boldsymbol{\mu}+\boldsymbol{h}^{(0)}(\bS_{i};\boldsymbol{\beta})+
G_i\left(\boldsymbol{\delta}+\boldsymbol{h}^{(1)}(\bS_{i};\boldsymbol{\beta})\right).
$$

\noindent Similar to the marginal case, we denote:

$$
\boldsymbol{h}^{(g)}(\bS_{i};\boldsymbol{\beta}) = (h_1^{(g)}(\bS_{i};\boldsymbol{\beta}),\cdots,h_T^{(g)}(\bS_{i};\boldsymbol{\beta}))^\top,
$$

\noindent where $\boldsymbol{h}^{(1)}$ is the treatment-specific surrogate component. Similar to the marginal model, let $\boldsymbol{\mu}^{*},\boldsymbol{\beta}^*,\boldsymbol{\delta}^{8}$ denote the pseudo-true parameters. 

By (\ref{problimit}), these quantities are defined as the solutions to the population score equations:
\begin{eqnarray}
\mathbb{E}[A\bepsilon_{i}^*]&=&\boldsymbol{0} \label{condeq1}\\
\mathbb{E}[J_{\boldsymbol{h}^{(0)}}(\bS_{i};\boldsymbol{\beta}^*)^{\top}A\bepsilon_{i}^* + G_iJ_{\boldsymbol{h}^{(1)}}(\bS_{i};\boldsymbol{\beta}^*)^{\top}A\bepsilon_{i}^*]&=&\boldsymbol{0} \label{condeq2}\\
\mathbb{E}[G_iA\bepsilon_{i}^*]&=&\boldsymbol{0} \label{condeq3}
\end{eqnarray}

\noindent {where here, $\bepsilon_{i}^* = \bY_i - \boldsymbol{\mu}^* - \boldsymbol{h}^{(0)}(\bS_{i};\boldsymbol{\beta}^*)-
G_i\left(\boldsymbol{\delta}^*+\boldsymbol{h}^{(1)}(\bS_{i};\boldsymbol{\beta}^*)\right).$} Equation (\ref{condeq1}) and (\ref{condeq3}) imply that:

$$
\mathbb{E}[A\bepsilon_{i}^*|G_i=0]=\boldsymbol{0},\quad
\mathbb{E}[A\bepsilon_{i}^*|G_i=1]=\boldsymbol{0}.
$$

\noindent It follows that:

$$
A\boldsymbol{\mu}^*=A\left(\mathbb{E}[\boldsymbol{Y}_{i}|G_i=0]-\mathbb{E}\left[\boldsymbol{h}^{(0)}(\bS_{i};\boldsymbol{\beta}^*)|G_i=0\right]\right),
$$
and
$$
\begin{aligned}
A\boldsymbol{\delta}^*&=A\left(\mathbb{E}[\bY_{i}|G_i=1]-\mathbb{E}\left[\boldsymbol{h}^{(0)}(\bS_{i};\boldsymbol{\beta}^*)+\boldsymbol{h}^{(1)}(\bS_{i};\boldsymbol{\beta}^*)\middle|G_i=1\right]\right)\\
&-A\left(\mathbb{E}[\bY_{i}|G_i=0]-\mathbb{E}\left[\boldsymbol{h}^{(0)}(\bS_{i};\boldsymbol{\beta}^*)\middle|G_i=0\right]\right).
\end{aligned}
$$

Now, by condition \textbf{(C4)}, $A$ is non-singular, thus:

$$
\begin{aligned}
\boldsymbol{\mu}^*&=\mathbb{E}[\boldsymbol{Y}_{i}|G_i=0]-\mathbb{E}\left[\boldsymbol{h}^{(0)}(\bS_{i};\boldsymbol{\beta}^*)|G_i=0\right],\\
\boldsymbol{\delta}^*&=\mathbb{E}[\bY_{i}|G_i=1]-\mathbb{E}\left[\boldsymbol{h}^{(0)}(\bS_{i};\boldsymbol{\beta}^*)+\boldsymbol{h}^{(1)}(\bS_{i};\boldsymbol{\beta}^*)\middle|G_i=1\right]\\
&-\mathbb{E}[\bY_{i}|G_i=0]+\mathbb{E}\left[\boldsymbol{h}^{(0)}(\bS_{i};\boldsymbol{\beta}^*)\middle|G_i=0\right].
\end{aligned}
$$

\noindent Now, define the true conditional mean as:
$$
q_{1}(\boldsymbol{s})=\mathbb{E}[\bY_{i}|\bS_{i}=\boldsymbol{s},G_i=1],
$$
and the pseudo-true working mean as:
$$
q_{1}^*(\boldsymbol{s})=\boldsymbol{\mu}^*+\boldsymbol{\delta}^*+\boldsymbol{h}^{(0)}(\boldsymbol{s};\boldsymbol{\beta}^*)+\boldsymbol{h}^{(1)}(\boldsymbol{s};\boldsymbol{\beta}^*).
$$

\noindent Under Assumptions \textbf{(D4)}, the residual treatment effect at time $t$ can be written as:

$$
\Delta_R(t)=\int q_{1,t}(\boldsymbol{s})dF_{\bS_t|G=0}(\boldsymbol{s})-\mathbb{E}[Y_{i,t}|G_i=0]
$$

\noindent and the corresponding pseudo-true limit of the estimator is:

$$
\Delta_R^*(t)
=
\int q_{1,t}^*(\boldsymbol{s})dF_{\bS_t|G=0}(\boldsymbol{s})
-
\mathbb{E}[Y_{i,t}|G_i=0].
$$

\noindent Using the expression for $\boldsymbol{\mu}^*$, this can be rewritten as:

$$
\Delta_R^*(t)=\delta_t^*+\mathbb{E}\left[h_t^{(1)}(\bS_{i,t};\beta_t^*)\middle| G_i=0\right].
$$

\noindent By replacing $\mathbb{E}\left[h_t^{(1)}(\bS_{i,t};\beta_t^*)\middle| G_i=0\right]$ with a Monte Carlo approximation, we obtain the formula for the PTE estimator in Subsection \ref{subsec:PTE}.

The bias induced by potential misspecification of the conditional mean is thus:

$$
B_R(t):=\Delta_R^*(t)-\Delta_R(t)
=
\int \left(q_{1,t}^*(\boldsymbol{s})-q_{1,t}(\boldsymbol{s})\right)dF_{\bS_t| G=0}(\boldsymbol{s}).
$$

\noindent Equivalently, if we define the residual as $e_t^*(\boldsymbol{s})=q_{1,t}(\boldsymbol{s})-q_{1,t}^*(\boldsymbol{s})$, then:

$$
B_R(t)
=
-\mathbb{E}\left[e_t^*(\bS_{i,t})| G_i=0\right].
$$

\noindent This identity shows that misspecification matters only through the expected error of the treated conditional mean under the \emph{control-group} surrogate distribution. In particular, $B_R(t)$ may be zero even when the working mean is
misspecified, provided that the misspecification residual integrates to zero under
$F_{\bS_t|G=0}$.

The resulting bias in the local and cumulative PTEs is:

$$
LPTE^*(t)-LPTE(t)=-\frac{B_R(t)}{\Delta(t)},
$$
and
$$
CPTE^*(t)-CPTE(t)=-\frac{\sum_{h=0}^t B_R(h)}{\sum_{h=0}^t \Delta(h)}.
$$

\noindent Thus, a positive bias in the residual treatment effect translates directly into a downward bias in the LPTE and CPTE.

Finally, the bias can be removed by construction if the treated model is fit under a criterion weighted toward the control-group surrogate distribution \citep{SHIMODAIRA2000227,Steingrimsson2022}. Specifically if the working model contains an intercept and the pseudo-true fit is defined using weights proportional to:
\begin{equation}
w_{i,t}(\boldsymbol{s}_{i,t})=\frac{f_{\bS_t|G=0}(\boldsymbol{s}_{i,t})}{f_{\bS_t|G=g_i}(\boldsymbol{s}_{i,t})}, \label{den_weight}
\end{equation}
where $w_{i,t}$ is the weight attributed to subject $i$ at time $t$, $\boldsymbol{s}_{i,t}$ is the \textbf{observed} surrogate trajectory to time $t$ for subject $i$ and $g_i$ is the observed treatment assignment for subject $i$, then the corresponding projection residual has mean zero under $F_{\bS_t|G=0}$, implying $\Delta_R^*(t)=\Delta_R(t)$ even under working-model misspecification. This can be achieved by estimating the distribution of the surrogate in both treatment arms. 


It is important to note that the  density in the numerator and denominator of (\ref{den_weight}) is the \textit{joint}  density of the surrogate trajectory. Ideally, the weights $w_{i,t}(\boldsymbol{s}_{i,t})$ would be estimated using a nonparametric procedure. However, nonparametric estimation of the numerator and denominator densities separately is generally impractical, particularly when $\boldsymbol{s}_{i,t}$ corresponds to a long surrogate trajectory. Fortunately, feasible alternative nonparametric approaches are available in the literature. For example, \cite{Sugiyama2008}, \cite{SUGIYAMA2011183} and \cite{Izbicki2014}  describe methods that estimate the weights $w_{i,t}(\boldsymbol{s}_{i,t})$ more directly, thereby avoiding the instability of separate density estimation. 

Finally, it is worth noting that the weights proposed here are very general. Above, we have described the weights in the context of the conditional model, but one could also consider such a bias-correction approach for the marginal model. For the marginal model with covariates, the bias-correcting weights would have the form:

$$
w_{i}(\boldsymbol{x}_i)=\frac{f_{\boldsymbol{X}_i|G=0}(\boldsymbol{x}_i)}{f_{\boldsymbol{X}_i|G=g_i}(\boldsymbol{x}_i)}.
$$

\noindent In this case, since the covariates are treated as fixed at baseline, the weights are the same overtime. Naturally, when the treatment is randomized, as is our setting, the weights simply become $1$, that is, the treatment-effect estimator from the marginal model is consistent by default.

For the covariates-adjusted conditional model, the weights become:

$$
w_{i,t}(\boldsymbol{x}_i,\boldsymbol{s}_i)=\frac{f_{\boldsymbol{X},\bS_t|G=0}(\boldsymbol{x}_i,\boldsymbol{s}_{i,t})}{f_{\boldsymbol{X},\bS_t|G=g_i}(\boldsymbol{x}_i,\boldsymbol{s}_{i,t})}.
$$

\noindent As noted above, these general results have been established in existing work. Our aim above was to describe the results in our setting. In our numerical studies, we do not implement this correction as our goal in those studies was to examine robustness under other types of misspecification.

\subsubsection*{Subject-level heteroskedastic model}

Note that a crucial aspect of the proofs presented so far is that $A$ is the same for all subjects. The matrix $A$ is implied by the chosen temporal dynamic of the model, and its being fixed implies that all subjects share the same temporal dynamic. This may be an unreasonable assumption in some scenarios, so it becomes worthwhile to develop the bias, and an appropriate bias correction, for the more general case. In this subsection, we show that the adapted bias-correcting weights for this case are equivalent to not using subject-specific weighting.

This case deserves its own separate treatment, as it requires a different set of arguments than the subject-level homoskedastic model. Indeed, as an example of how this case differs from the previous ones, an essential step in deriving the bias, or lack thereof, for the marginal model is noting that, by randomization:

$$
\mathbb{E}[\boldsymbol{g}(\bX_{i};\boldsymbol{\beta}^*)|G_i=1]=\mathbb{E}[\boldsymbol{g}(\bX_{i};\boldsymbol{\beta}^*)]
$$

\noindent In the heteroskedastic case, we would actually need that:

$$
\mathbb{E}[A_i\boldsymbol{g}(\bX_{i};\boldsymbol{\beta}^*)|G_i=1]=\mathbb{E}[A_i\boldsymbol{g}(\bX_{i};\boldsymbol{\beta}^*)],
$$

\noindent If $A_i$ is defined using post-treatment information or quantities affected by treatment, then randomization does not suffice to guarantee that the above equality holds. This breaks the remainder of the argument.

In this subsection, we continue to assume that the weight matrix is known, but now allow each individual to have their own weight matrix, denoted by $A_i$. Unlike in the homoskedastic case, the extension to settings in which $A_i$ is estimated is not straightforward. Nevertheless, by showing that the bias correction cancels the subject-level weighting in this simplified, idealized setting, we illustrate why the two should generally not be used together. Lastly, in this case, the calculation for the marginal and conditional models do not differ in a meaningful way. Thus, we develop the bias correction for the conditional model, but the same development can also be used for the marginal model by omitting $\boldsymbol{S}_i$.

Let $\bZ_i = (\bX_i^{\top},\bS_i^{\top})^{\top}$, with $\bZ_{i,t} = (\bX_i^{\top},\bS_{i,t}^{\top})^{\top}$. We assume the following working mean:

$$
\boldsymbol{m}(\bZ_{i},G_i;\boldsymbol{\mu},\boldsymbol{\beta},\boldsymbol{\delta})=\boldsymbol{\mu}+\boldsymbol{g}^{(0)}(\bZ_{i};\boldsymbol{\beta})+
G_i\left(\boldsymbol{\delta}+\boldsymbol{g}^{(1)}(\bZ_{i};\boldsymbol{\beta})\right).
$$

\noindent The score equation implies that the pseudo-true values $(\boldsymbol{\mu}^*,\boldsymbol{\beta}^*,\boldsymbol{\delta}^*)$ solve:
\begin{eqnarray}
\mathbb{E}\left[A_i\bepsilon_i^*\right]&=&\boldsymbol{0},\label{eq4}\\
\mathbb{E}\left[J_{g^{(0)}}(\bZ_{i};\boldsymbol{\beta}^*)A_i\bepsilon_i^*+J_{g^{(1)}}G_i(\bZ_{i};\boldsymbol{\beta}^*)A_i\bepsilon_i^*\right]&=&\boldsymbol{0},  \label{eq5}\\
\quad\mathbb{E}\left[G_iA_i\bepsilon_i^*\right]&=&\boldsymbol{0}  \label{eq6}.
\end{eqnarray}

\noindent Equations (\ref{eq4}) and (\ref{eq6}) implies:

\begin{equation}
\begin{aligned}
\mathbb{E}[A_i] \boldsymbol{\mu}^{*}=&\mathbb{E}\left[A_i\bY_{i}\right]-\mathbb{E}\left[A_i\boldsymbol{g}^{(0)}(\bZ_{i};\boldsymbol{\beta}^*)\right]\\
&-\rho\left(\mathbb{E}[A_i|G_i=1]\boldsymbol{\delta}^*+\mathbb{E}[A_i\boldsymbol{g}^{(1)}(\bZ_{i};\boldsymbol{\beta}^*)|G_i=1]\right),\label{eq8}
\end{aligned}
\end{equation}

\begin{equation}
\begin{aligned}
\mathbb{E}[A_i|G_i=1]\boldsymbol{\delta}^*&=\mathbb{E}\left[A_i\bY_{i}|G_i=1\right]
-\mathbb{E}\left[A_i|G_i=1\right]\boldsymbol{\mu}^*\\
&-\mathbb{E}\left[A_i\left(\boldsymbol{g}^{(0)}(\bZ_{i};\boldsymbol{\beta}^*)+\boldsymbol{g}^{(1)}(\bZ_{i};\boldsymbol{\beta}^*)\right)\middle|G_i=1\right].
\end{aligned}\label{eq9}
\end{equation}

\noindent From there, we can note that $\mathbb{E}[A_i] = \rho\mathbb{E}[A_i|G_i=1] + (1-\rho)\mathbb{E}[A_i|G_i=0]$, thus, from Equation (\ref{eq8}) we have:

\begin{equation}
\begin{aligned}
\mathbb{E}[A_i|G_i=1] \boldsymbol{\mu}^{*}=&\frac{1}{\rho}\left(\mathbb{E}\left[A_i\bY_{i}\right]-\mathbb{E}\left[A_i\boldsymbol{g}^{(0)}(\bZ_{i};\boldsymbol{\beta}^*)\right]\right)\\
&-\left(\mathbb{E}[A_i|G_i=1]\boldsymbol{\delta}^*+\mathbb{E}[A_i\boldsymbol{g}^{(1)}(\bZ_{i};\boldsymbol{\beta}^*)|G_i=1]\right)\\
& - \frac{(1-\rho)}{\rho}\mathbb{E}[A_i|G_i=0]\boldsymbol{\mu}^{*}. 
 \label{eq8implies}
\end{aligned}
\end{equation}

We can repeat a similar argument to break this down further:

\begin{equation}
\begin{aligned}
\mathbb{E}[A_i|G_i=1] \boldsymbol{\mu}^{*}&=\mathbb{E}\left[A_i\bY_{i}\middle|G_i=1\right]- \mathbb{E}[A_i|G_i=1]\boldsymbol{\delta}^*\\
&-\mathbb{E}\left[A_i\left(\boldsymbol{g}^{(0)}(\bZ_{i};\boldsymbol{\beta}^*)+\boldsymbol{g}^{(1)}(\bZ_{i};\boldsymbol{\beta}^*)\right)\middle|G_i=1\right] \\
&+\frac{1-\rho}{\rho}\bigg(\mathbb{E}\left[A_i\bY_{i}\middle|G_i=0\right]\\
&-\mathbb{E}\left[A_i\boldsymbol{g}^{(0)}(\bZ_{i};\boldsymbol{\beta}^*)\middle|G_i=0\right] - \mathbb{E}[A_i|G_i=0]\boldsymbol{\mu}^{*}\bigg)\label{eq7ImpliesAgain}
\end{aligned}
\end{equation}

Bring this back to equation \ref{eq9}, we have that:

$$
\begin{aligned}
\mathbb{E}[A_i|G_i=1]\boldsymbol{\delta}^*&= \mathbb{E}[A_i|G_i=1]\boldsymbol{\delta}^*-\frac{1-\rho}{\rho}\bigg(\mathbb{E}\left[A_i\bY_{i}\middle|G_i=0\right]\\
&-\mathbb{E}\left[A_i\boldsymbol{g}^{(0)}(\bZ_{i};\boldsymbol{\beta}^*)\middle|G_i=0\right] - \mathbb{E}[A_i|G_i=0]\boldsymbol{\mu}^{*}\bigg)
\end{aligned}
$$

From this we can conclude that:

$$
\begin{aligned}
\mathbb{E}\left[A_i\bY_{i}\middle|G_i=0\right]-\mathbb{E}\left[A_i\boldsymbol{g}^{(0)}(\bZ_{i};\boldsymbol{\beta}^*)\middle|G_i=0\right] - \mathbb{E}[A_i|G_i=0]\boldsymbol{\mu}^{*}=\boldsymbol{0}
\end{aligned}
$$

Which, along with equation \ref{eq4}, allow us to conclude that:

$$
\begin{aligned}
\boldsymbol{\mu}^{*}&=\mathbb{E}[A_i|G_i=0]^{-1}\mathbb{E}\left[A_i\bY_{i}\middle|G_i=0\right]-\mathbb{E}\left[A_i\boldsymbol{g}^{(0)}(\bZ_{i};\boldsymbol{\beta}^*)\middle|G_i=0\right]\\
\boldsymbol{\delta}^{*} &= \mathbb{E}[A_i|G_i=1]^{-1}\mathbb{E}\left[A_i\bY_{i}\middle|G_i=1\right]- \boldsymbol{\mu}^{*}\\
 &- \mathbb{E}[A_i|G_i=1]^{-1}\left(\mathbb{E}\left[A_i\left(\boldsymbol{g}^{(0)}(\bZ_{i};\boldsymbol{\beta}^*)+\boldsymbol{g}^{(1)}(\bZ_{i};\boldsymbol{\beta}^*)\right)\middle|G_i=1\right]\right)
\end{aligned}
$$

Similar to the conditional case, we define the true conditional mean
$$
q_{1}(\boldsymbol{z})=\mathbb{E}[\bY_{i}|\bZ_{i}=\boldsymbol{z},G_i=1],
$$
and the pseudo-true working mean
$$
\begin{aligned}
q_{1}^*(\boldsymbol{z})&=\boldsymbol{\mu}^*+\boldsymbol{\delta}^*+\boldsymbol{g}^{(0)}(\boldsymbol{z};\boldsymbol{\beta}^*)+\boldsymbol{g}^{(1)}(\boldsymbol{z};\boldsymbol{\beta}^*)\\
&=\cancel{\boldsymbol{\mu}^*}+\boldsymbol{g}^{(0)}(\boldsymbol{z};\boldsymbol{\beta}^*)+\boldsymbol{g}^{(1)}(\boldsymbol{z};\boldsymbol{\beta}^*)\\
&+\mathbb{E}[A_i|G_i=1]^{-1}\mathbb{E}\left[A_i\bY_{i}\middle|G_i=1\right]- \cancel{\boldsymbol{\mu}^{*}}\\
&-\mathbb{E}[A_i|G_i=1]^{-1}\left(\mathbb{E}\left[A_i\left(\boldsymbol{g}^{(0)}(\bZ_{i};\boldsymbol{\beta}^*)+\boldsymbol{g}^{(1)}(\bZ_{i};\boldsymbol{\beta}^*)\right)\middle|G_i=1\right]\right)\\
&=\mathbb{E}[A_i|G_i=1]^{-1}\mathbb{E}\left[A_i\bY_{i}\middle|G_i=1\right]\\
&+\boldsymbol{g}^{(0)}(\boldsymbol{z};\boldsymbol{\beta}^*)-\mathbb{E}[A_i|G_i=1]^{-1}\mathbb{E}\left[A_i\boldsymbol{g}^{(0)}(\bZ_{i};\boldsymbol{\beta}^*)\middle|G_i=1\right]\\
&+\boldsymbol{g}^{(1)}(\boldsymbol{z};\boldsymbol{\beta}^*)-\mathbb{E}[A_i|G_i=1]^{-1}\mathbb{E}\left[A_i\boldsymbol{g}^{(1)}(\bZ_{i};\boldsymbol{\beta}^*)\middle|G_i=1\right]
\end{aligned}
$$

\noindent Under Assumptions \textbf{(D4)}, the residual treatment effect at time $t$ can be written as:

$$
\Delta_R(t)=\int q_{1,t}(\boldsymbol{z})dF_{\bZ_t|G=0}(\boldsymbol{s})-\mathbb{E}[Y_{i,t}|G_i=0]
$$

\noindent and the corresponding pseudo-true limit of estimator is:

$$
\Delta_R^*(t)
=
\int q_{1,t}^*(\boldsymbol{z})dF_{\bZ_t|G=0}(\boldsymbol{z})
-
\mathbb{E}[Y_{i,t}|G_i=0].
$$

We therefore end up in a situation similar to that of the conditional model in the homoskedastic case, where the bias induced by misspecification of the conditional mean is:

$$
B_R(t):=\Delta_R^*(t)-\Delta_R(t)
=
\int q_{1,t}^*(\boldsymbol{z})-q_{1,t}(\boldsymbol{z}) dF_{\bZ_t| G=0}(\boldsymbol{z}).
$$

Similarly, if we define the residual as $e_t^*(\boldsymbol{z})=q_{1,t}(\boldsymbol{z})-q_{1,t}^*(\boldsymbol{z})$, then:

$$
B_R(t)
=
-\mathbb{E}\left[e_t^*(\bZ_{i,t})| G_i=0\right].
$$

To correct this bias, as before, we can apply a bias correction that corresponds to changing the score equation so as to ensure that $\mathbb{E}\left[e_t^*(\bZ_{i,t}) \mid G_i=0\right] = 0$. The equivalent bias-corrected weights then take the form:

$$
A_i^{\text{bc}}=\left(W_i A_i^{-1}\right)A_i = W_i,
$$

\noindent where $W_i = \text{diag}\{w_{i,1},\cdots,w_{i,T}\}$ are the weights for subject $i$ according to the previous bias-correction.

It is straightforward to see, by substituting these weights into the score equation, that they guarantee $B_R(t) = 0$. However, doing so is also equivalent to not using subject-specific weights. Thus, this approach offers no clear advantage over using the same weight matrix $A$ for all subjects, and may in fact be less desirable.

One alternative would be to use different weights that are not defined at the subject level. This can work as long as the distribution of the weights is the same in the control and treatment groups. Otherwise, correcting the bias due to model misspecification will also undo the intended weighting. Alternatively, if the model is correctly specified, then no bias correction is needed, and there is no issue with using subject-level weights.

Finally, we emphasize that, when the working mean model is correctly specified, the choice between global and subject-level temporal dynamics affects only model efficiency. In contrast, when the working mean model is misspecified, only the global temporal-dynamics model may still guarantee a consistent estimator of the PTE. Therefore, using a global temporal-dynamics model is a valid and conservative choice, as it permits the use of the bias correction described above. Subject-level temporal dynamics may improve efficiency when they more accurately reflect the data-generating process, but this potential gain comes with the risk of introducing bias.

}

\subsection{Lingering Effects \label{app:lingering}}
Here, we present sufficient conditions for $\PTE^{*} \le \PTE$, which is equivalent to $\Delta^{*}_{R,\text{Area}} \ge \Delta_{R,\text{area}}$. Without loss of generality, in order to simplify notation, assume that the dependence is only up to the previous time. That is, suppose that only the $t$ and $t-1$ surrogate values had an affect on the outcome at time $t$, such that

$$
\begin{aligned}
\Delta_{R,\text{area}} 
&= \sum_{t=1}^{T}\int \mathbb{E}[Y^{(1)}_t|S^{(1)}_{t}=S^{(0)}_{t}=s_t,S^{(1)}_{t-1}=S^{(0)}_{t-1}=s_{t-1}]dF_{S\supzero_t,S\supzero_{t-1}}(s_{t-1},s_t) \\
&-\int \mathbb{E}[Y^{(0)}_t] |S^{(1)}_{t}=S^{(0)}_{t}=s_t,S^{(1)}_{t-1}=S^{(0)}_{t-1}=s_{t-1}]dF_{S\supzero_t,S\supzero_{t-1}}(s_{t-1},s_t)\\
&= \sum_{t=1}^{T}\int \int \mathbb{E}[Y^{(1)}_t|S^{(1)}_{t}=s_t,S^{(1)}_{t-1}=s_{t-1}]dF_{S\supzero_{t-1}|S\supzero_{t}=s_{t}}(s_{t-1})dF_{S\supzero_{t}}(s_{t})\\ 
&-\mathbb{E}[Y^{(0)}_t].
\end{aligned}
$$

It follows that 

$$
\begin{aligned}
\Delta_{R,\text{area}}^{*}&= \sum_{t=1}^{T}\int\mathbb{E}[Y^{(1)}_t -Y^{(0)}_t|S^{(1)}_{t}=S^{(0)}_{t}=s_t]dF_{S^{(0)}_t}(s_t)\\
&= \sum_{t=1}^{T}\int\mathbb{E}[Y^{(1)}_t|S^{(1)}_{t}=s_t]dF_{S^{(0)}_t}(s_t)- \sum_{t=1}^{T}\int\mathbb{E}[Y^{(0)}_t|S^{(0)}_{t}=s_t]dF_{S^{(0)}_t}(s_t)\\
&= \sum_{t=1}^{T}\int\int\mathbb{E}[Y^{(1)}_t|S^{(1)}_{t}=s_t,S^{(1)}_{t-1}=s_{t-1}]dF_{S^{(1)}_{t-1}|S_t^{(1)}=s_t}(s_{t-1})dF_{S^{(0)}_t}(s_t)\\ 
&-\mathbb{E}[Y^{(0)}_t].
\end{aligned}
$$

and thus, we can focus on comparing 

$$
\int\mathbb{E}[Y^{(1)}_t|S^{(1)}_{t}=s_t,S^{(1)}_{t-1}=s_{t-1}]dF_{S^{(0)}_{t-1}|S_t^{(0)}=s_t}(s_{t-1})
$$

to

$$
\int\mathbb{E}[Y^{(1)}_t|S^{(1)}_{t}=s_t,S^{(1)}_{t-1}=s_{t-1}]dF_{S^{(1)}_{t-1}|S_t^{(1)}=s_t}(s_{t-1}).
$$

The only difference between these two expressions is the integration measure,
$F_{S^{(1)}_{t-1}|S^{(1)}_t=s_t}$ versus
$F_{S^{(0)}_{t-1}|S^{(0)}_t=s_t}$.
Although Condition \textbf{(S3)} imposes forward stochastic dominance,
it does not, by itself, imply any ordering of the conditional distributions of $S_{t-1}$ given $S_t$.
That is, forward stochastic dominance of $S_t$ given past history does not generally translate into backward dominance of $S_{t-1}$ given the current surrogate value. Thus, to make a statement comparing these measures, we impose a backward stochastic dominance condition formulated conditional on the shared surrogate history up to time $t-2$. If we do not condition on the shared surrogate history up to time $t-2$, a definitive statement about the difference between these quantities cannot be made, as it cannot be characterized without additional assumptions. Thus, if the following condition holds, $\PTE^{*} \le \PTE$:
\begin{itemize}
    \item[\textbf{(S4)}] \textbf{Backwards positive treatment effect on the surrogate effect}:
    $\mathbb{P}(S^{(1)}_{t-1}>s_{t-1}|S^{(1)}_{t}=s_t,\bS_{t-2}\supone=\boldsymbol{s})\ge \mathbb{P}(S^{(0)}_{t-1}>s_{t-1}|S^{(0)}_{t}=s_t,\bS_{t-2}\supzero=\boldsymbol{s}),$ for all trajectories $\boldsymbol{s}$, all $s_{t-1}$ and $s_t$ and for all times $t$.
\end{itemize}

\noindent Under \textbf{(S4)}, the resulting truncated estimator of the PTE is guaranteed to be conservative, in the sense that it underestimates the true PTE.

\clearpage 
\section{} 

\subsection{SSM, OLS, and LMM \label{app:ssmolslmm}} 

Here, we discuss how the SSM approach is similar to and differs from ordinary least squares (OLS) and a linear mixed model (LMM). A natural comparator to a SSM is the LMM, which can also be viewed as a special case of Generalized Least Squares (GLS). By iteratively substituting the state equation (\ref{eq.2}) into the observation equation (\ref{eq.1}), the SSM can be written as an LMM with the innovations ${\bomega_t}$ as independent random effects and a block lower-triangular, highly structured design matrix. In that sense, there is some equivalence between both approaches.

In practical terms, the main differences among the SSM, OLS, and LMM approaches lie in the dependence structure each one implies, since all of them are special cases of GLS. In this sense, OLS is equivalent to assuming independence among measurements from the same patient. While both the SSM and LMM account for within-subject dependence, they do so in different ways. Specifying $W_t$ via a discount strategy produces smooth, time-varying evolution variances; between-patient heterogeneity can be introduced by allowing patient-specific $W_{i,t}$. In contrast, standard LMMs are often fit with simpler variance structures by default, though they can be extended to allow both time- and group-specific heteroskedasticity.

In practice, both time- and group-specific heteroskedasticity are plausible. For example, if $Y_t$ is blood pressure, older patients may exhibit larger evolution variances than younger patients, while the variance may be roughly constant over time. Conversely, if $Y_t$ measures allergy symptoms, certain seasons may have higher variance than others, potentially overshadowing between-patient differences. Ideally, we would like to know each individual's time-specific variance, but that is simply unreasonable. Estimating the variance will inevitably require some structural assumption. The SSMs discount strategy imposes such structure, allowing individual- and time-specific variances, but it supports only smooth changes over time; for short series, estimates of $W_t$ will be nearly constant.

Ultimately, relative performance depends on which model better reflects the data-generating process. Because the SSM is expressly built for temporal data, taking into account its directional evolution, we expect it to more closely match the true covariance structure in many settings. Beyond statistical efficiency, the SSM also has computational benefits: its Markovian structure enables Kalman filtering and smoothing, yielding substantially more efficient fitting than solving the equivalent mixed model directly.

\subsection{Discount Strategy \label{app:discount}}

The model proposed in Section \ref{sec:estimating} is a set of independent random walks for each individual mean, with a shared treatment effects $\delta_{g,t}, g=1,2$. Because it effectively acts as a flexible filter or smoother of the trajectories, and is thus capable of representing arbitrary patterns, it can accommodate virtually any observed outcome trajectory.

The term $m_{i,t}$ is introduced to account for within-subject correlation over time, while $\mu_t$ captures the overall temporal trend in the outcome under the control group. Notably, the pairs $(\mu_t, m_{i,t})$ and $(\nu_t, n_{i,t})$ are not separately identifiable. However, the sums $\mu_t + m_{i,t}$ and $\nu_t + n_{i,t}$ are identifiable and estimable from the observed data, and these combined quantities are sufficient for defining and estimating our proposed treatment effect and PTE quantities. Although we do not impose such a constraint here, one possible way to address the lack of separate identifiability would be to normalize the subject-specific effects by requiring $\sum_i m_{i,t} = 0$ for each $t$.

The evolution covariance matrices $W_{\mu,t}$, $W_{i,t}$, $W'_{\nu,t}$ and $W'_{i,t}$ are specified using a discount strategy, which is typical in SSMs. First, a set of discount factors $d_{\mu,t}, d_{i,t}, d'_{\nu,t}, d'_{i,t} \in (0,1]$ is selected, and the evolution covariance matrix $W_{i,t}$ is expressed as: 

$$
W_{i,t} = \frac{1-d_{i,t}}{d_{i,t}} C_{i,t-1},
$$
where $C_{i,t-1}$ represents the filtered variance for the estimator of $m_{i,t-1}$, as obtained using the Kalman filter, and analogous formulas are used for the other covariance matrices. The choice of discount factors does not materially affect the point estimator of $\btheta_{i,t}$, provided the factor is strictly less than $1$ (to avoid a null covariance matrix). That said, the discount factor does affect the estimator’s efficiency: smaller values yield more flexible estimates, but with higher variance. 

At this point, it is useful to discuss the interpretation of the discount factor. A discount factor of $d$ can be viewed as the proportion of information retained from one time point to the next, so that $1-d$ is the proportion of information lost at each step. Thus, if $d=1$ there is no information loss (knowledge about $\btheta_{i,t-1}$ carries over without degradation to $\btheta_{i,t}$), whereas if $d\approx 0$, $\btheta_{i,t-1}$ and $\btheta_{i,t}$ are nearly unrelated. In this sense, if we believe there is a $10\%$ loss of information over one year, the corresponding discount factor would be about $0.9^{1/2}\approx 0.948$ for semiannual measurements and $0.9^{1/365}\approx 0.9997$ for daily measurements. Previous studies can provide insight into the percentage of information lost over a given period, which can then serve as a basis for choosing a discount factor in the current study, as in a random walk model (like the one we are proposing), the discount factor is equivalent to the square of the correlation between two adjacent times. In doing so, it is preferable to slightly undershoot the implied discount factor, as this amounts to erring on the side of a more flexible specification.

As noted in \cite{West-DLM}, it is common to select values between $0.9$ and $0.99$, starting with larger values and then comparing several options to determine which performs best. In the context of surrogate evaluation, we advise using a smaller value, as this is a more conservative choice, although how small depends on the nature and frequency of the data. For instance, $0.9$ may be too large for data measured every six months, while being too small for data measured daily.


To obtain the estimator for $\btheta_{t}$, we can set $V_t^{(g)}$ as $1$, as, when using a discount strategy for the evolution matrix, changes in the scale of $V_t^{(g)}$ do not affect the point estimate. They do affect the uncertainty quantification, but since we use a nonparametric bootstrap to obtain variances and confidence intervals (described in \ref{app:boot}), we can set $V_t^{(g)}$ to an arbitrary value. That being said, it can be useful to allow for $V_t^{(g)}$ to depend on $t$ or $g$. In those cases, one can use the methods proposed in Sections 10.7 and 10.8 of \cite{West-DLM} or the method proposed in \cite{Alves-kparametric}, all of which can account for both phenomena and integrate easily with the proposed bootstrap strategy.

\subsection{Local PTE and the Cumulative PTE Estimators \label{app:estimate}}

In Section \ref{sec:estimating}, we provided the explicit form of the proposed estimator of the $\PTE = \CPTE(T)$. Here, we provide the explicit parallel plug-in form of the proposed estimators of the Local and Cumulative PTE under the SSM specification in the main text. By the derivations in the main text, the local treatment and residual treatment effects admit the estimators
$$
\widehat\Delta(t)=\widehat\delta_{2,t},
\qquad
\widehat\Delta_R(t)=\widehat\delta_{1,t}+\frac{1}{n_0}\sum_{G_i=0}\Big(\widehat f_t^{(1)}(\bS_{i,t})-\widehat f_t^{(0)}(\bS_{i,t})\Big),
$$
and therefore the local PTE estimator is
$$
\widehat{\LPTE}(t)=1-\frac{\widehat\Delta_R(t)}{\widehat\Delta(t)}
=
1-\frac{\widehat\delta_{1,t}
+\frac{1}{n_0}\sum_{G_i=0}\Big(\widehat f_t^{(1)}(\bS_{i,t})-\widehat f_t^{(0)}(\bS_{i,t})\Big)}
{\widehat\delta_{2,t}}.
$$
The cumulative PTE estimator up to time $t$ is
$$
\widehat{\CPTE}(t)=1-\frac{\sum_{h=0}^{t}\widehat\Delta_R(h)}{\sum_{h=0}^{t}\widehat\Delta(h)}
=
1-\frac{\sum_{h=0}^{t}\widehat\delta_{1,h}
+\sum_{h=0}^{t}\frac{1}{n_0}\sum_{G_i=0}\Big(\widehat f_h^{(1)}(\bS_{i,h})-\widehat f_h^{(0)}(\bS_{i,h})\Big)}
{\sum_{h=0}^{t}\widehat\delta_{2,h}}
$$
where $\widehat{\delta}_1$, $\widehat{\delta}_2$, $\widehat{f}_{t}^{(1)}(\bS_{i,t})$ and $\widehat{f}_{t}^{(1)}(\bS_{i,t})$ are obtained using the Kalman Filter and Smoother algorithms.

\subsection{Nonparametric Bootstrap Procedure \label{app:boot}}

Here, we provide details and justification for our proposed nonparametric bootstrap procedure. To develop this strategy, we use the fact that, for linear Gaussian SSMs, the Kalman filter coincides with the Bayesian posterior mean. This identity allows us to decompose the estimator into per-patient contributions, which can be computed once and then recombined at each bootstrap iteration. We emphasize that we are not adopting a Bayesian inferential framework (though one could); our uncertainty quantification is frequentist. The Bayesian connection is used purely as an algebraic device to facilitate the bootstrap. 

In a Bayesian framework, we seek the posterior distribution of the latent states $\btheta_t$, with the posterior mean typically serving as the point estimator. Under the model specifications in (1) and (2) of section \ref{sec:estimating} and a Gaussian prior $\btheta_1 \sim \mathcal{N}(a_1, R_1)$, the posterior distribution of $\btheta_t$ is also Gaussian and its means and covariances can be computed efficiently via the Kalman filter and smoother \citep{Kalman_filter_origins}. State–space models have the following notable properties:
\begin{itemize}
    \item \textbf{Conditional independence.} The observations $\{Y_t\}_{t=1}^{T}$ are independent given the latent states $\{\btheta_t\}_{t=1}^{T}$. In particular, any dependence between $Y_i$ and $Y_j$ arises solely through the dependence between $\btheta_i$ and $\btheta_j$.
    \item \textbf{Markov property.} The latent process is first–order Markov: $\btheta_t$ depends on $\btheta_{t-k}$ only through $\btheta_{t-1}$. Consequently, the observations inherit this dependence structure via the states.
\end{itemize}
\noindent These are properties of the model rather than requirements of the true data–generating process. We explicitly state them here because they are leveraged for our bootstrap strategy, although only as algebraic properties of the estimator.

We describe our procedure for a general $N$-variate time series governed by Equations (\ref{eq.1})–(\ref{eq.2}) from section \ref{sec:estimating}, $\{\bY_t\}_{t=1}^{T}$, with a Gaussian prior $\btheta_1 \sim \mathcal{N}(a_1, R_1)$. We express $\bY_t = (Y_{1,t},\dots,Y_{N,t})^{\top} \in \mathbb{R}^{N}$ and use a diagonal covariance matrix for the observational model. We do \textbf{not} assume that this specification is correct, nor an accurate representation of the data generating process. This is used only to highlight the algebraic properties of the estimator. 
Let $\pi$ denote the prior/posterior distribution of the full state trajectory $\btheta = \{\btheta_t\}_{t=1}^{T}$. Accordingly, $\pi(\cdot |\mathcal{D}_0)$ denotes the prior distribution on $\btheta$, and $\pi(\cdot |\mathcal{D}_T)$ denotes the smoothed posterior given the full data $\mathcal{D}_T = \{\bY_1,\dots,\bY_T\}$. Using  Bayes' rule it follows that:

$$
\begin{aligned}
\pi(\btheta|\mathcal{D}_T) &\propto \pi(\btheta|\mathcal{D}_0) f(Y_1,\dots,Y_t|\btheta)= \pi(\btheta|\mathcal{D}_0) \prod_{t=1}^{T} f(Y_t|\btheta,Y_1,\dots,Y_{t-1})\\
\end{aligned}
$$

\noindent Now, using the conditional independence and the Markov property of state-space models, we have that:

$$
\begin{aligned}
\pi(\btheta|\mathcal{D}_T) &\propto  \pi(\btheta|\mathcal{D}_0) \prod_{t=1}^{T} f(Y_t|\btheta,Y_1,\dots,Y_{t-1}) =  \pi(\btheta|\mathcal{D}_0) \prod_{t=1}^{T} f(Y_t|\btheta_t\\
&=  \pi(\btheta|\mathcal{D}_0) \prod_{t=1}^{T} \prod_{i=1}^{N} f(Y_{i,t}|\btheta_t)  =    \prod_{i=1}^{N}\pi(\btheta|\mathcal{D}_0)^{\frac{1}{N}}\prod_{t=1}^{T}  f(Y_{i,t}|\btheta_t)\\
\end{aligned}
$$

\noindent Let us call $p_i(\btheta)=\pi(\btheta|\mathcal{D}_0)^{1/N}\prod_{t=1}^{T}  f(Y_{i,t}|\btheta_t)$. Note that we can interpret $p_i$ as the posterior for a SSM where we only observed the $i$th replication of the experiment. In particular, $p_i$ is the density for a Gaussian distribution whose mean and variance can be obtained using the Kalman Filter and Smoother. Let $m_{i,t}$ and $C_{i,t}$ be the mean and covariance matrix of $p_i$. Since $\pi(\cdot|\mathcal{D}_T)$ is a product of Gaussians, we know that it is also Gaussian with mean and covariance matrix defined by:

$$
m_t=C_t\left(\sum_{i=1}^{N}C_{i,t}^{-1}m_{i,t}\right), \quad C_{t} = \left(\sum_{i=1}^{N}C_{i,t}^{-1}\right)^{-1}.
$$

\noindent Notably, this formulation does not help us in terms of calculating $m_t$ and $C_t$ in general, as it would have the same computational cost as the Kalman Filter itself, but it will be useful for our bootstrap approach. Specifically, let $h_j(i)$ be the $i$th index from the $j$th iteration of the bootstrap. It follows then that:

$$
m_t^{(j)}=C_t^{(j)}\left(\sum_{i=1}^{N}C_{h_j(i),t}^{-1}m_{h_j(i),t}\right), \quad C_{t}^{(j)} = \left(\sum_{i=1}^{N}C_{h_j(i),t}^{-1}\right)^{-1}.
$$

\noindent Thus, we can compute $m_{i,t}$ and $C_{i,t}$ once, and then average them to obtain each iteration of the bootstrap, making the cost of running the bootstrap with $B$ replications to be $\mathcal{O}(NT) + \mathcal{O}(BN)$, instead of $\mathcal{O}(BNT)$. 


Lastly, we discuss the conditions under which this bootstrap strategy provides a valid approximation to the sampling distributions of the treatment effects and the PTE. Note that these are not conditions for the estimators themselves to be valid, such as \textbf{(C1)--(C6)}, nor are they conditions for the causal interpretation to be valid, such as \textbf{(S1)--(S3)}. Rather, these conditions are only required for the uncertainty quantification based on this specific bootstrap approach to be representative of the true uncertainty. If these conditions are violated, the estimators and their causal interpretation remain valid, provided that \textbf{(C1)--(C6)} and \textbf{(S1)--(S3)} hold. In that case, the implication is that a different approach to uncertainty quantification would be needed. First, for this bootstrap strategy to provide a valid approximation to the sampling distributions of the treatment effects, we require the following condition:

\textbf{Subject-level i.i.d. units:} \textit{The individual trajectories $(\mathbf{Y}_i,\mathbf{S}_i)$ are independent and identically distributed across $i$. That is, the joint distribution of the outcome and surrogate trajectories is the same for all individuals, and individuals are mutually independent.}

\noindent Next, for the distribution of the PTE estimator, we also require:

\textbf{PTE is well-defined:} \textit{Beyond \textbf{(C1)}, the total treatment effect satisfies $\sum_{t=0}^{T}\Delta(t) \ge \epsilon$ for some $\epsilon>0$, so that the PTE is well-defined and nondegenerate.}

\noindent Lastly, we also require \textbf{joint resampling for joint estimators}. That is, to capture the dependence between the marginal and conditional estimators when forming the PTE, we must use the same bootstrap index set for both models, yielding paired draws $\big(\widehat{\Delta}_R^{(b)}(t),\widehat{\Delta}^{(b)}(t)\big)$. This is not a condition on the data, but rather a fundamental part of the bootstrap procedure.

In general, some regularity conditions are also required for estimators to which the bootstrap is applied. However, for our method, these are satisfied without further assumptions, since $\widehat{\Delta}(t)$ and $\widehat{\Delta}_R(t)$ are sample-mean contrasts, and the PTE is a continuously differentiable function of those means with a nonzero denominator (by the second condition above). Hence, the subject-level bootstrap is valid for these particular estimators.

\clearpage
\section{} 
\subsection{Simulation Settings: Data Generation \label{app:sims}}
\allowdisplaybreaks

For all settings, our general data generating model had the following specification: 
\begin{align*}
Y_{i,t} &= \mu_{i,t} + \beta S_{i,t}  + \gamma_1 H_1(t) G_{i},\\
\mu_{i,t} &= \phi_1\mu_{i,t-1} + \text{sign}\times V_{i,t}\times\omega_{1,i,t},\\
S_{i,t} &= \nu_{i,t} + \gamma_2 H_2(t)G_{i},\\
\nu_{i,t} &= \phi_2\nu_{i,t-1} + \omega_{2,i,t},\\
\omega_{1,i,t} &\sim \mathcal{LG}(\alpha,\exp(\psi(\alpha))),\\
\omega_{2,i,t} &\sim \mathcal{N}(0,W),\\
\frac{\mu_{i,1}}{\sqrt{V_{i,1}/(1-\phi_2)}} &\sim \mathcal{LG}(\alpha,\exp(\psi(\alpha))),\\
\nu_{i,1} &\sim \mathcal{N}(0,W/(1-\phi_2)),\\
V_{i,t} &\sim \mathcal{G}(\tau/2,\tau/2V)
\end{align*}

\noindent where we will detail each component in the following text. It is reasonable to question why such data generation was used. Essentially, our aim was to generate data such that the  distributions of the observational and evolution errors are purposefully non-symmetric, heavy-tailed, and involve nonlinear temporal dynamics. That is, complicated distributions that would be difficult to correctly specify in practice without true knowledge of the data generating approach. We now walk through the specific quantities in the specification above: \begin{itemize}
    \item  $H_1(t)$ and $H_2(t)$ are the two treatment effect functions where $H_1(t)$ represents the direct effect of the treatment on $Y_{i,t}$, and $H_2(t)$ represents the effect of the treatment on the surrogate. The parameters $\gamma_1$ and $\gamma_2$ control the scale. These are the functions that we varied across our settings to create different treatment effect and PTE trajectories. The specific way in which we varied the functions is described further below. 
    \item $\mathcal{LG}(\alpha,\beta)$ represents a Log-Gamma distribution with parameters $\alpha$ and $\beta$ such that $\beta= \exp(\psi(\alpha))$, and $\psi$ is the \textit{digamma} function. By setting $\beta= \exp(\psi(\alpha))$, we have that $\mathbb{E}[\omega_{1,i,t}]=0$. The shape parameter $\alpha$ controls how asymmetric the distribution of $\omega_{1,i,t}$ is, with larger values representing a symmetric distribution that is close to a Gaussian. For this study, we use $\alpha=1$.
    \item The $\text{sign}$ variable indicates the side of the asymmetry. The Log-Gamma is left skewed, so when sign is $1$ we have left skewed innovations, and when $\text{sign}$ is $-1$ we have right skewed innovations.
    \item The auxiliary scale $V_{i,t}$ is generated from a Gamma distribution to emulate a $t$-like tail. Specifically, when $\tau/2 \gg 1$, then $ V_{i,t}\times\omega_{1,i,t}$ has a distribution close to a Student-$t$ with $ \tau$ degrees of freedom. For this study, we use $\tau=10$ and $V=0.0025$.
    \item The variance of the state-evolution innovations is specified relative to the observational error variance. Specifically, we set $W=0.2 \times V$.
\end{itemize}

In our examination of the proposed PTE estimator and corresponding confidence interval, goals (1) and (2) in section \ref{sec:simul}, we specify $H_1(t)$ and $H_2(t)$ such that the treatment effect changes over time, but the PTE itself is constant. Specifically, we consider three distinct behaviors for the treatment effect: monotonically increasing, increasing and then decreasing (no effect in the beginning and the end of the study) and a random walk. Figure \ref{fig:Ht} illustrates the pattern in each case.  Within each treatment effect behavior type, we additionally vary $H_1(t)$ and $H_2(t)$ to produce varying PTE levels, specifically PTEs of 0.25, 0.5, 0.75, 0.9, and 1.0. This results in a total of 15 different settings. Within each, we also examine results across the following sample sizes: 50, 100, 150, 200, 250, and 300.

In our examination of the proposed test for temporal homogeneity, goal (3) in section \ref{sec:simul}, we specify $H_1(t)$ and $H_2(t)$ such that the PTE does or does not change over time (depending on the setting). Specifically, we examine the following five different scenarios, illustrated in Figure \ref{fig:simul7_scenarios}, where each represents a different dynamic for the PTE:

\begin{enumerate}
    \item \textbf{Scenario 1 (null hypothesis):} The treatment effect is monotone increasing, as in the monotone setting for the first set of simulations, and the PTE does \textit{not} change over time.
    \item \textbf{Scenario 2 (Good surrogate when treatment is weak):} The PTE is close to $1$ when the treatment effect is small and close to $0$ when the treatment effect is large. Depending on the duration of the study, one may conclude that the surrogate is good, when in fact, it is not. 
    \item \textbf{Scenarios 3 (Good surrogate when treatment is strong):} This scenario is essentially the inverse case of scenario 2.
    \item \textbf{Increasing PTE:} In this case, by the end of the study the surrogate is valid, but due to the initial low PTE, the cumulative PTE is below 0.75.
    \item \textbf{Decreasing PTE:} The initial PTE is large indicating a valid surrogate, but substantially weakens by the end of the study. The cumulative PTE is above $0.75$ at the end of the study. 
\end{enumerate}

Scenarios $2$ and $3$ are meant to represent treatments with seasonal effects, such as a treatment for an allergy. Scenario $2$ represents the case where the effect captured by the surrogate is constant and negligible, while the treatment itself is useful, but only in certain periods (the seasons where the individual does have the symptoms). Scenario $3$ is the opposite: the direct effect of the treatment is negligible and constant, while the effect captured by the surrogate is large, but only in certain periods.

Scenarios $4$ and $5$ are meant to represent treatments for which the effect takes some time to ``kick-in''. Scenario $4$ has the direct effect increase over time, while the effect captured by the surrogate is small and negligible. Scenario $5$ represents the opposite.

Within each scenario we examine results across the following sample sizes: 50, 100, 150, 200, 250, and 300; and across the following durations: 1, 2, 3, 5, and 10 years.

Furthermore, our implementation deliberately avoided using any knowledge of the true data-generating mechanism; we simply fit the generic model introduced in section \ref{sec:simul}. For simplicity, we excluded covariates, omit interactions between the surrogate and the treatment, and impose a linear effect for the surrogate. These choices are made solely for expository clarity: our framework is not restricted to these assumptions and can be modified to account for such effects.


\subsection{Additional Simulation Results: Surrogate Evaluation \label{app:sims2}}

Bias results in the main text show that our proposed method has consistently smaller bias than the other methods. Figure \ref{fig:simul3_coverage} shows coverage levels close to the nominal level of 0.95 for the proposed method in all settings, and over-coverage for the LMM and GEE in some settings.  Figure \ref{fig:simul3_ic_width} shows the average width of the confidence intervals, and indicates that the proposed method yields considerably tighter confidence intervals, with the OLS and GEE approaches showing the worst performance in most cases. With respect to surrogate validity testing, the main text illustrates the simulation results in a figure; in Table \ref{tab:simul3_hyp_test_end}, we additionally provide these results in table form.

\subsection{Additional Simulation Results: Temporal Homogeneity \label{app:sims3}}

The main text illustrates these simulation results in a figure; in Tables \ref{tab:simul7_case1} through \ref{tab:simul7_case5}, we additionally provide these results in table form.

\clearpage
\section{Additional Analyses of the Diabetes Trial Application \label{app:diabetes}}

Our primary analysis in section \ref{sec:app} used a model with up to 28 surrogate lags (7 years) and included two binary baseline covariates (sex and smoking status); our model specification was linear and additive. We additionally evaluated several specifications for how the outcome depends on surrogate history, including the use of splines and interaction terms that allowed the surrogate effect to differ by treatment arm. Results from these various models did not meaningfully differ from the estimated PTE when using the linear model. However, these various models added many more parameters that limited the number of surrogate lags we could include without instability and increased uncertainty. For these reasons, we focused on results using the linear specification for the relationship between surrogate and outcome and use the opportunity to expand more (below) on the impact of lags on the results.

In section \ref{sec:app}, our focus was on making a decision about the strength of the surrogate. We additionally highlighted how explicitly modeling the temporal relationship between the surrogate and outcome by including lingering effects of past surrogate values can substantially change the estimated PTE. Specifically, we presented results using 28 lags versus 0 lags. Here, we explore lingering effects in more detail by examining other potential lag choices. 

For each maximum lag $K$ (from $K=0$ up to $K=37$), we fit a conditional model including $(S_t,S_{t-1},\ldots,S_{t-K})$ and computed the local PTE and cumulative PTE over time. We included two binary baseline covariates (sex and smoking status), as in the main text.  In Figure \ref{fig:applied1a} we show the estimated CPTE trajectory for the subset of models with no lags (also in the main text), a maximum of 4 lags (1 year), a maximum of 12 lags (3 years), a maximum of 20 lags (5 years), and a maximum of 28 lags (7 years, also in the main text). Pointwise confidence intervals are not shown for clarity. Point estimates and confidence intervals for the CPTE at the last time $T$ using these varying lags are shown in Table \ref{tab:applied_cpte_end}. 

Two features are apparent in these results. First, CPTE tends to increase as more surrogate history is incorporated.  This pattern is consistent with the presence of lingering effects: past HbA1c values contain additional information about subsequent AER beyond what is captured by the contemporaneous HbA1c alone. Second, CPTE trajectories show some early-time variability, which is expected because the treatment effect is relatively small at the beginning of follow-up, making the PTE inherently noisier; this early variability has limited influence on the end-of-study CPTE.

Figure \ref{fig:applied1b} presents the CPTE at the final observed time as a function of $K$. As $K$ increases, the point estimate rises significantly up to  $K = 30$, but uncertainty grows substantially and the point estimate begins to decline slightly for the largest lags. This late-lag behavior should be interpreted cautiously: lags close to $T$ are only observed for participants who entered early and remained under observation through the end of follow-up, and thus, the effective sample size for estimating the contribution of the most distant lag terms is small. Consequently, CPTE estimates for $K$ close to $T$ have much higher uncertainty.


In addition, we applied the proposed temporal homogeneity test using the fitted models and found that the test result was sensitive to the lag length $K$: for small to moderate $K$, conclusions switched between rejecting and not rejecting homogeneity, while for larger $K$ the p-values tend to be larger, consistently not rejecting homogeneity. This pattern is consistent with two nonexclusive explanations: (i) when lag history is truncated, residual temporal dependence can appear as time-variation in the PTE; and (ii) increasing $K$ may inflate uncertainty, reducing test power.

It is important to emphasize that the CPTE estimates, as well as the conclusion of the time-homogeneity test, may differ across choices of the maximum lag, but none of them is ``wrong" per se. Rather, as discussed in Section \ref{subsec:lingering}, changing the maximum lag changes the target estimand. 
Specifically, the CPTE computed with maximum lag $K$ corresponds to the proportion of the treatment effect explained by the most recent $K$ surrogate measurements, which is generally distinct from the proportion explained by the full surrogate trajectory. Thus, differences across $K$ should be interpreted as differences in the estimand being evaluated, not as evidence that the estimator is invalid. Moreover, the truncated estimand can still be informative about the full-history PTE and, under additional conditions, provides a conservative lower bound for it.

More concretely, if lingering effects are modeled only up to lag $K<T$, then for times $t \le K$ the estimated PTE up to time $t$ coincides with the ``complete'' PTE: the treatment was not applied on times before $0$, so the treatment effect cannot be captured by the surrogate at those times and omitting them does not change the estimand target. In contrast, for times $t > K$, part of the treatment effect may be captured by lags beyond $K$; omitting those pathways changes the estimand, so the target PTE for $t>K$ may not be equal to the complete PTE. Consequently, even if the complete PTE is constant over time, a truncated lag specification can induce apparent time variation because the impact of omitting lingering effects differs across times.

Importantly, we do not present these analyses as formal hypothesis tests with multiple testing adjustments; rather, they are intended to illustrate how conclusions can change when lingering surrogate effects are omitted. In settings where a decision hinges on
rejecting or not rejecting homogeneity, we do \textit{not} suggest running separate tests for each model specification. Instead, a single maximum lag for the surrogate should be selected a priori, guided by the effective sample size at each lag (as we did in the main text), before evaluating the PTE. Taken together, these additional results illustrate a central contribution of our framework: in a jointly longitudinal setting, surrogate evaluation can be affected by the inclusion/exclusion of lingering effects of the surrogate on the outcome. In particular, incorporating a richer surrogate history can reveal a substantially larger PTE, by accounting for lingering effects that may be difficult to detect or accommodate in approaches that do not explicitly model the outcome as a time-series.

\clearpage
\bibliography{Surrogate_bib}

@article{parast2021hetero,
  title={Testing for heterogeneity in the utility of a surrogate marker},
  author={Parast, Layla and Cai, Tianxi and Tian, Lu},
  journal={Biometrics},
  volume={79},
  number={2},
  pages={799--810},
  year={2023},
  publisher={The International Biometric Society}
}

@article{elliott2023surrogate,
	author = {Elliott, Michael R},
	journal = {Annual Review of Statistics and its Application},
	pages = {75--96},
	publisher = {Annual Reviews},
	title = {Surrogate Endpoints in Clinical Trials},
	volume = {10},
	year = {2023}}

@article{elliott2015surrogacy,
  title={Surrogacy marker paradox measures in meta-analytic settings},
  author={Elliott, Michael R and Conlon, Anna SC and Li, Yun and Kaciroti, Nico and Taylor, Jeremy MG},
  journal={Biostatistics},
  volume={16},
  number={2},
  pages={400--412},
  year={2015},
  publisher={Oxford University Press}
}

@article{chen2007criteria,
  title={Criteria for surrogate end points},
  author={Chen, Hua and Geng, Zhi and Jia, Jinzhu},
  journal={Journal of the Royal Statistical Society Series B: Statistical Methodology},
  volume={69},
  number={5},
  pages={919--932},
  year={2007},
  publisher={Oxford University Press}
}

@article{lin1997estimating,
	Author = {Lin, DY and Fleming, TR and De Gruttola, V and others},
	Journal = {Statistics in medicine},
	Number = {13},
	Pages = {1515--1527},
	Title = {Estimating the proportion of treatment effect explained by a surrogate marker},
	Volume = {16},
	Year = {1997}}

@article{wang2002measure,
	Author = {Wang, Yue and Taylor, Jeremy MG},
	Journal = {Biometrics},
	Number = {4},
	Pages = {803--812},
	Publisher = {Wiley Online Library},
	Title = {A measure of the proportion of treatment effect explained by a surrogate marker},
	Volume = {58},
	Year = {2002}}

@article{vanderweele2013surrogate,
  title={Surrogate measures and consistent surrogates},
  author={VanderWeele, Tyler J},
  journal={Biometrics},
  volume={69},
  number={3},
  pages={561--565},
  year={2013},
  publisher={Wiley Online Library}
}

@article{blette2023low,
  title={Is low-risk status a surrogate outcome in pulmonary arterial hypertension? An analysis of three randomised trials},
  author={Blette, Bryan S and Moutchia, Jude and Al-Naamani, Nadine and Ventetuolo, Corey E and Cheng, Chao and Appleby, Dina and Urbanowicz, Ryan J and Fritz, Jason and Mazurek, Jeremy A and Li, Fan and others},
  journal={The Lancet Respiratory Medicine},
  volume={11},
  number={10},
  pages={873--882},
  year={2023},
  publisher={Elsevier}
}

@article{hsiao2025avoiding,
  title={Avoiding the surrogate paradox: an empirical framework for assessing assumptions},
  author={Hsiao, Emily and Tian, Lu and Parast, Layla},
  journal={Journal of Nonparametric Statistics},
  pages={1--22},
  year={2025},
  publisher={Taylor \& Francis}
}

@article{hsiao2025resilience,
  title={Resilience Measures for the Surrogate Paradox},
  author={Hsiao, Emily and Tian, Lu and Parast, Layla},
  journal={arXiv preprint arXiv:2506.12194},
  year={2025}
}

@article{diabetes1993effect,
  title={Diabetes Control and Complications Trial Research Group: The effect of intensive treatment of diabetes on the development and progression of long-term complications in insulin-dependent diabetes mellitus},
  author={{D}{C}{C}{T}},
  journal={New England journal of medicine},
  volume={329},
  number={14},
  pages={977--986},
  year={1993},
  publisher={Mass Medical Soc}
}

@ARTICLE{Arulampalam-particlefilter,
  author={Arulampalam, M.S. and Maskell, S. and Gordon, N. and Clapp, T.},
  journal={IEEE Transactions on Signal Processing}, 
  title={A tutorial on particle filters for online nonlinear/non-Gaussian Bayesian tracking}, 
  year={2002},
  volume={50},
  number={2},
  pages={174-188},
  doi={10.1109/78.978374}}

@misc{Alves-kparametric,
      title={k-parametric Dynamic Generalized Linear Models: a sequential approach via Information Geometry},
      author={Mariane Branco Alves and Helio S. Migon and Raíra Marotta and Silvaneo V. dos Santos Jr.},
      year={2024},
      eprint={2201.05387},
      archivePrefix={arXiv},
      primaryClass={stat.ME}
}

@book{West-DLM,
  author = {West, Mike and Harrison, Jeff},
  howpublished = {Hardcover},
  isbn = {0387947256},
  month = {February},
  opturl = {http://www.amazon.fr/exec/obidos/ASIN/0387947256/citeulike04-21},
  publisher = {Springer-Verlag},
  title = {Bayesian Forecasting and Dynamic Models (Springer Series in Statistics)},
  year = 1997
}

@article{WestHarrMigon,
author = { Mike   West  and  P.   Jeff   Harrison  and  Helio S.   Migon },
title = {Dynamic Generalized Linear Models and Bayesian Forecasting},
journal = {Journal of the American Statistical Association},
volume = {80},
number = {389},
pages = {73-83},
year  = {1985},
publisher = {Taylor \& Francis}

}

@article{Kalman_filter_origins,
    Author = {Kalman, Rudolph Emil},
    Title = {A New Approach to Linear Filtering and Prediction Problems},
    Journal = {Transactions of the ASME--Journal of Basic Engineering},
    Volume = {82},
    Number = {Series D},
    Pages = {35--45},
    Year = {1960}
}

@techreport{Athey2019Index,
 title = "The Surrogate Index: Combining Short-Term Proxies to Estimate Long-Term Treatment Effects More Rapidly and Precisely",
 author = "Athey, Susan and Chetty, Raj and Imbens, Guido W and Kang, Hyunseung",
 institution = "National Bureau of Economic Research",
 type = "Working Paper",
 series = "Working Paper Series",
 number = "26463",
 year = "2019",
 month = "November",
 doi = {10.3386/w26463},
}

@article{Chetty2014Education,
Author = {Chetty, Raj and Friedman, John N. and Rockoff, Jonah E.},
Title = {Measuring the Impacts of Teachers II: Teacher Value-Added and Student Outcomes in Adulthood},
Journal = {American Economic Review},
Volume = {104},
Number = {9},
Year = {2014},
Month = {September},
Pages = {2633–79}
}

@article{Roberts2022Longitudinal,
    author = {Roberts, Emily K. and Elliott, Michael R. and Taylor, Jeremy M. G.},
    title = {Solutions for Surrogacy Validation with Longitudinal Outcomes for a Gene Therapy},
    journal = {Biometrics},
    volume = {79},
    number = {3},
    pages = {1840-1852},
    year = {2022},
    month = {07}
}

@article{Agniel2024Survival,
    author = {Agniel, Denis and Parast, Layla},
    title = {Robust evaluation of longitudinal surrogate markers with censored data},
    journal = {Journal of the Royal Statistical Society Series B: Statistical Methodology},
    volume = {87},
    number = {3},
    pages = {891-907},
    year = {2024},
    month = {12}
}

@article{Agniel2020Longitudinal,
    author = {Agniel, Denis and Parast, Layla},
    title = {Evaluation of Longitudinal Surrogate Markers},
    journal = {Biometrics},
    volume = {77},
    number = {2},
    pages = {477-489},
    year = {2020},
    month = {06},
    issn = {0006-341X},
    doi = {10.1111/biom.13310},
}

@article{Diaz2023Mediation,
title = {Efficient and flexible mediation analysis with time-varying mediators, treatments, and confounders},
author = {Iván Díaz and Nicholas Williams and Kara E. Rudolph},
pages = {20220077},
volume = {11},
number = {1},
journal = {Journal of Causal Inference},
doi = {doi:10.1515/jci-2022-0077},
year = {2023},
lastchecked = {2025-09-19}
}

@article{Alves2025Econometrics,
author = {Alves, Mariane B. and Migon, Helio S. and Menezes, André F. B. and Pinheiro, Eduardo G. and dos Santos Jr., Silvaneo V.},
title = {Dynamic Econometric Models: A State-Space Formulation},
journal = {Journal of Forecasting},
volume = {n/a},
number = {n/a},
pages = {},
year = 2025
}

@article{chakraborty2014dynamic,
  title={Dynamic treatment regimes},
  author={Chakraborty, Bibhas and Murphy, Susan A},
  journal={Annual review of statistics and its application},
  volume={1},
  number={1},
  pages={447--464},
  year={2014},
  publisher={Annual Reviews}
}

@article{Prashad2025Epidemiology,
title = {State-space modelling for infectious disease surveillance data: Stochastic simulation techniques and structural change detection},
journal = {Infectious Disease Modelling},
volume = {10},
number = {4},
pages = {1507-1532},
year = {2025},
issn = {2468-0427},
author = {Christopher D. Prashad}
}

@article{SHIMODAIRA2000227,
title = {Improving predictive inference under covariate shift by weighting the log-likelihood function},
journal = {Journal of Statistical Planning and Inference},
volume = {90},
number = {2},
pages = {227-244},
year = {2000},
issn = {0378-3758},
author = {Hidetoshi Shimodaira}
}

@article{Steingrimsson2022,
    author = {Steingrimsson, Jon A and Gatsonis, Constantine and Li, Bing and Dahabreh, Issa J},
    title = {Transporting a Prediction Model for Use in a New Target Population},
    journal = {American Journal of Epidemiology},
    volume = {192},
    number = {2},
    pages = {296-304},
    year = {2022},
    month = {07},
    issn = {0002-9262},
    doi = {10.1093/aje/kwac128},
}

@article{Wager2016,
author = {Stefan Wager  and Wenfei Du  and Jonathan Taylor  and Robert J. Tibshirani },
title = {High-dimensional regression adjustments in randomized experiments},
journal = {Proceedings of the National Academy of Sciences},
volume = {113},
number = {45},
pages = {12673-12678},
year = {2016},
doi = {10.1073/pnas.1614732113}}

@article{Sugiyama2008,
  author  = {Sugiyama, Masashi and Suzuki, Taiji and Nakajima, Shinichi and Kashima, Hisashi and von B{\"u}nau, Paul and Kawanabe, Motoaki},
  title   = {Direct importance estimation for covariate shift adaptation},
  journal = {Annals of the Institute of Statistical Mathematics},
  year    = {2008},
  volume  = {60},
  number  = {4},
  pages   = {699--746},
  doi     = {10.1007/s10463-008-0197-x},
}

@article{SUGIYAMA2011183,
title = {Direct density-ratio estimation with dimensionality reduction via least-squares hetero-distributional subspace search},
journal = {Neural Networks},
volume = {24},
number = {2},
pages = {183-198},
year = {2011},
issn = {0893-6080},
author = {Masashi Sugiyama and Makoto Yamada and Paul {von Bünau} and Taiji Suzuki and Takafumi Kanamori and Motoaki Kawanabe}
}

@inproceedings{Izbicki2014,
author = {Izbicki, Rafael and Lee, Ann and Schafer, Chad},
year = {2014},
month = {04},
title = {High-Dimensional Density Ratio Estimation with Extensions to Approximate Likelihood Computation},
journal = {Proceedings of the International Conference on Artificial Intelligence and Statistics, AISTATS}
}

\clearpage 
\begin{figure}
    \centering
    \includegraphics[scale=1]{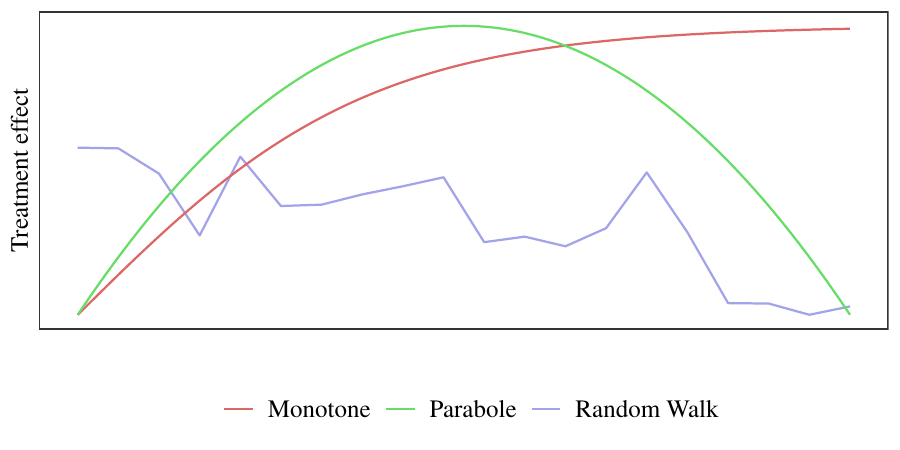}
    \caption{Varying treatment effect trajectories examined in the simulation study: monotone (red), parabola (green) and random walk (blue).}
    \label{fig:Ht}
\end{figure}

\clearpage 
\begin{figure}
\centering
    \begin{subfigure}[c]{0.19\textwidth}
    \includegraphics[width=1\linewidth]{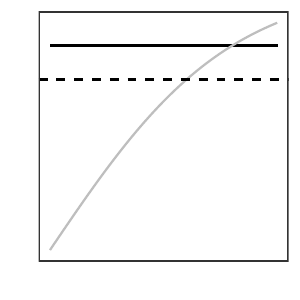}
    \caption{Scenario 1.}\label{fig:simul7_scenario_1}
    \end{subfigure}
    \begin{subfigure}[c]{0.19\textwidth}
    \includegraphics[width=1\linewidth]{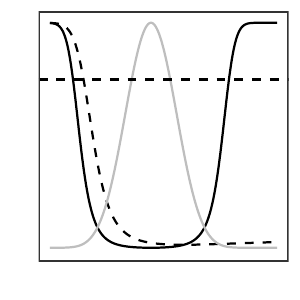}
    \caption{Scenario 2.}\label{fig:simul7_scenario_2}
    \end{subfigure}
    \begin{subfigure}[c]{0.19\textwidth}
     \centering
    \includegraphics[width=1\linewidth]{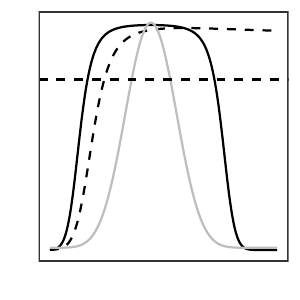}
    \caption{Scenario 3.}\label{fig:simul7_scenario_3}
    \end{subfigure}
    \begin{subfigure}[c]{0.19\textwidth}
     \centering
    \includegraphics[width=1\linewidth]{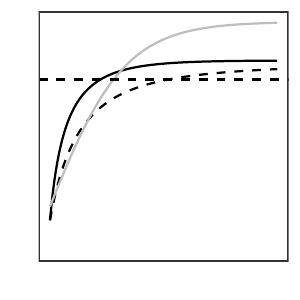}
    \caption{Scenario 4.}\label{fig:simul7_scenario_4}
    \end{subfigure}
    \begin{subfigure}[c]{0.19\textwidth}
     \centering
    \includegraphics[width=1\linewidth]{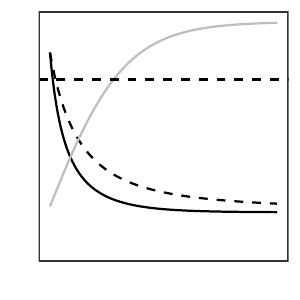}
    \caption{Scenario 5.}\label{fig:simul7_scenario_5}
    \end{subfigure}
    \caption{Varying trajectories examined in the simulation study assessing performance of the proposed temporal homogeneity test: Local PTE (solid black line), cumulative PTE (dashed black line) and total (direct plus indirect) local treatment effect (solid grey line) for each scenario; the horizontal dashed line represents the 0.75 threshold.}
    \label{fig:simul7_scenarios}
\end{figure}

\clearpage
\begin{figure}
    \centering
    \includegraphics[scale=1]{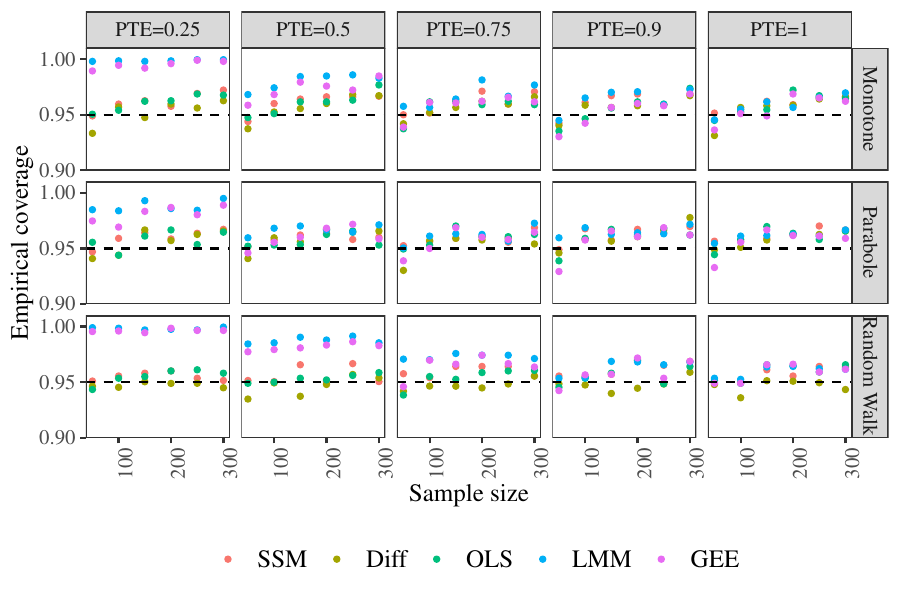}   
    \caption{Simulation results examining the empirical coverage of the constructed 95\% confidence intervals for the PTE using the proposed SSM estimator (red), compared to estimation using GEE (green), LMM (blue), Diff (olive) and OLS (purple), across 15 simulations which vary with respect to the treatment effect trajectory (monotone, parabole, or random walk) and the PTE (0.25, 0.5,0.75,0.9, or 1.0), and 6 sample sizes; a horizontal black dashed line is shown at the nominal level of 0.95.}
    \label{fig:simul3_coverage}
\end{figure}

\begin{figure}
    \centering
    \includegraphics[scale=1]{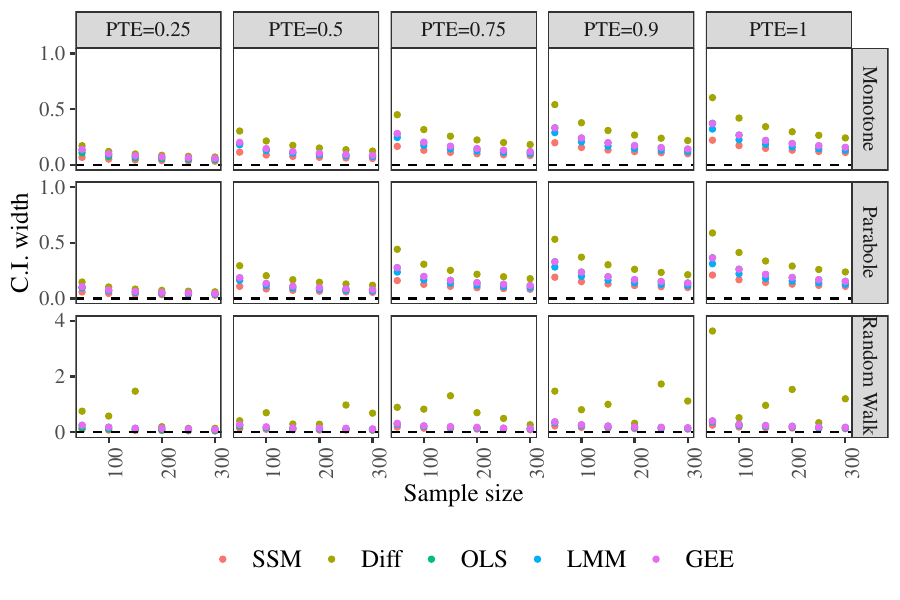}   
    \caption{Simulation results examining the average width of the constructed 95\% confidence intervals for the PTE using the proposed SSM estimator (red), compared to estimation using GEE (green), LMM (blue), Diff (olive) and OLS (purple), across 15 simulations which vary with respect to the treatment effect trajectory (monotone, parabole, or random walk) and the PTE (0.25, 0.5,0.75,0.9, or 1.0), and 6 sample sizes; a horizontal black dashed line is shown at a width of 0 for reference.}
    \label{fig:simul3_ic_width}
\end{figure}

\clearpage
\begin{table}
\centering
\begin{tabular}[t]{llrrrrr}
\hline
\multirow{2}*{Treatment} &  \multirow{2}*{Method} & \multicolumn{5}{c}  {PTE}\\
 &  & 0.25 & 0.5 & 0.75 & 0.9 & 1\\
\hline
 \multirow{4}*{Monotone} & Diff & 0.00\% & 0.00\% & 5.38\% & 62.24\% & 84.14\%\\
 & GEE & 0.00\% & 0.00\% & 5.39\% & 85.74\% & 96.25\%\\
 & LMM & 0.00\% & 0.00\% & 4.74\% & 91.32\% & 98.26\%\\
 & OLS & 0.00\% & 0.00\% & 5.50\% & 86.01\% & 96.38\%\\
 & SSM & 0.00\% & 0.00\% & 5.35\% & 97.95\% & 99.92\%\\
\hline
 \multirow{5}*{Parabole}   & Diff & 0.00\% & 0.00\% & 5.81\% & 63.56\% & 84.46\%\\
 & GEE & 0.00\% & 0.00\% & 5.40\% & 86.85\% & 96.75\%\\
 & LMM & 0.00\% & 0.00\% & 5.54\% & 92.18\% & 98.60\%\\
 & OLS & 0.00\% & 0.00\% & 5.41\% & 86.75\% & 96.77\%\\
 & SSM & 0.00\% & 0.00\% & 5.91\% & 98.49\% & 99.96\%\\
\hline
 \multirow{5}*{Random Walk}  & Diff & 0.00\% & 0.01\% & 4.41\% & 50.95\% & 71.10\%\\
 & GEE & 0.00\% & 0.00\% & 4.46\% & 82.01\% & 95.51\%\\
 & LMM & 0.00\% & 0.00\% & 3.74\% & 87.53\% & 97.49\%\\
 & OLS & 0.00\% & 0.00\% & 5.30\% & 84.03\% & 96.00\%\\
 & SSM & 0.00\% & 0.00\% & 5.17\% & 96.79\% & 99.87\%\\
\hline
 \multirow{5}*{Overall} & Diff & 0.00\% & 0.00\% & 5.20\% & 58.92\% & 79.91\%\\
 & GEE & 0.00\% & 0.00\% & 5.09\% & 84.87\% & 96.17\%\\
 & LMM & 0.00\% & 0.00\% & 4.67\% & 90.34\% & 98.12\%\\
 & OLS & 0.00\% & 0.00\% & 5.40\% & 85.60\% & 96.38\%\\
 & SSM & 0.00\% & 0.00\% & 5.48\% & 97.74\% & 99.91\%\\
\hline
\end{tabular}
\caption{Simulation results examining the proportion of rejections of the null hypothesis that the PTE is $\leq 0.75$ using the proposed SSM estimator, compared to estimation using GEE, LMM, Diff, and OLS, across 15 simulations which vary with respect to the treatment effect trajectory (monotone, parabole, or random walk) and the PTE (0.25, 0.5,0.75,0.9, or 1.0), and 6 sample sizes; note that when the PTE $=$ 0.25 or 0.5 or 0.75, the null hypothesis is true and thus this calculation proportion is the empirical Type 1 error, with PTE$=0.75$ reflecting the boundary of the null hypothesis, and when the PTE = 0.9 or 1.0, the null hypothesis is false and thus, this proportion reflects the empirical power.}
\label{tab:simul3_hyp_test_end}
\end{table}

\clearpage

\begin{table}
\centering
\resizebox{\textwidth}{!}{
\begin{tabular}{r @{\hskip 0.5cm} ll @{\hskip 0.5cm} ll @{\hskip 0.5cm} ll @{\hskip 0.5cm} ll @{\hskip 0.5cm} ll}
\toprule
Sample &  \multicolumn{2}{c}{1 Year} &  \multicolumn{2}{c}{2 Year} &  \multicolumn{2}{c}{3 Year} &  \multicolumn{2}{c}{5 Year} &  \multicolumn{2}{c}{10 Year}\\
size & \textbf{MSD} & \textbf{Wald} & \textbf{MSD} & \textbf{Wald} & \textbf{MSD} & \textbf{Wald} & \textbf{MSD} &\textbf{Wald} & \textbf{MSD} & \textbf{Wald}\\
\midrule
50 & 6.19\% & 4.50\% & 7.43\% & 15.45\% & 6.78\% & 37.28\% & 9.80\% & 92.03\% & 11.63\% & 100.00\%\\
100 & 4.80\% & 3.47\% & 6.39\% & 8.22\% & 6.78\% & 13.91\% & 6.58\% & 39.16\% & 8.56\% & 99.75\%\\
150 & 4.90\% & 3.66\% & 5.35\% & 5.40\% & 5.00\% & 9.65\% & 6.09\% & 19.46\% & 6.78\% & 85.20\%\\
200 & 5.00\% & 2.67\% & 5.25\% & 5.15\% & 6.14\% & 7.57\% & 5.74\% & 16.19\% & 6.73\% & 63.07\%\\
250 & 3.91\% & 2.28\% & 5.89\% & 5.05\% & 4.85\% & 5.25\% & 6.34\% & 11.93\% & 6.14\% & 46.88\%\\
300 & 5.54\% & 2.92\% & 6.44\% & 5.10\% & 5.69\% & 5.84\% & 5.25\% & 10.05\% & 5.54\% & 36.58\%\\
\bottomrule
\end{tabular}}
\caption{Simulation results examining the proportion of rejections of the null hypothesis that the PTE is constant over time i.e., temporal homogeneity, for Scenario 1 (Null hypothesis is true) at an $\alpha$-level of 0.05}
\label{tab:simul7_case1}
\end{table}

\begin{table}
\centering
\resizebox{\textwidth}{!}{
\begin{tabular}{r @{\hskip 0.5cm} ll @{\hskip 0.5cm} ll @{\hskip 0.5cm} ll @{\hskip 0.5cm} ll @{\hskip 0.5cm} ll}
\toprule
Sample &  \multicolumn{2}{c}{1 Year} &  \multicolumn{2}{c}{2 Year} &  \multicolumn{2}{c}{3 Year} &  \multicolumn{2}{c}{5 Year} &  \multicolumn{2}{c}{10 Year}\\
size & \textbf{MSD} & \textbf{Wald} & \textbf{MSD} & \textbf{Wald} & \textbf{MSD} & \textbf{Wald} & \textbf{MSD} &\textbf{Wald} & \textbf{MSD} & \textbf{Wald}\\
\midrule
50 & 17.32\% & 12.12\% & 18.87\% & 18.91\% & 21.95\% & 32.52\% & 28.44\% & 60.79\% & 44.37\% & 98.97\%\\
100 & 37.12\% & 27.65\% & 40.86\% & 34.50\% & 43.71\% & 46.95\% & 55.56\% & 70.36\% & 70.36\% & 99.21\%\\
150 & 53.94\% & 40.66\% & 60.43\% & 50.60\% & 61.39\% & 59.27\% & 72.62\% & 78.71\% & 84.40\% & 98.54\%\\
200 & 66.32\% & 52.15\% & 72.91\% & 59.87\% & 73.28\% & 66.69\% & 82.98\% & 81.29\% & 92.09\% & 98.44\%\\
250 & 75.17\% & 59.57\% & 81.75\% & 66.82\% & 81.99\% & 75.36\% & 90.03\% & 86.79\% & 95.76\% & 98.61\%\\
300 & 81.82\% & 68.31\% & 87.25\% & 74.90\% & 87.72\% & 80.33\% & 93.84\% & 88.81\% & 98.25\% & 98.71\%\\
\bottomrule
\end{tabular}}
\caption{Simulation results examining the proportion of rejections of the null hypothesis that the PTE is constant over time i.e., temporal homogeneity, for Scenario 2 (Null hypothesis is false) at an $\alpha$-level of 0.05}
\label{tab:simul7_case2}
\end{table}

\begin{table}
\centering
\resizebox{\textwidth}{!}{
\begin{tabular}{r @{\hskip 0.5cm} ll @{\hskip 0.5cm} ll @{\hskip 0.5cm} ll @{\hskip 0.5cm} ll @{\hskip 0.5cm} ll}
\toprule
Sample &  \multicolumn{2}{c}{1 Year} &  \multicolumn{2}{c}{2 Year} &  \multicolumn{2}{c}{3 Year} &  \multicolumn{2}{c}{5 Year} &  \multicolumn{2}{c}{10 Year}\\
size & \textbf{MSD} & \textbf{Wald} & \textbf{MSD} & \textbf{Wald} & \textbf{MSD} & \textbf{Wald} & \textbf{MSD} &\textbf{Wald} & \textbf{MSD} & \textbf{Wald}\\
\midrule
50 & 18.08\% & 14.24\% & 20.73\% & 22.42\% & 23.21\% & 34.54\% & 29.57\% & 61.92\% & 42.78\% & 98.74\%\\
100 & 36.09\% & 28.15\% & 41.72\% & 38.38\% & 44.11\% & 49.24\% & 53.51\% & 71.09\% & 69.97\% & 98.84\%\\
150 & 52.88\% & 40.63\% & 58.97\% & 49.50\% & 58.44\% & 57.75\% & 70.73\% & 76.13\% & 82.75\% & 98.51\%\\
200 & 63.08\% & 49.17\% & 69.80\% & 58.01\% & 70.56\% & 66.16\% & 80.86\% & 80.73\% & 89.54\% & 98.58\%\\
250 & 71.36\% & 57.32\% & 76.75\% & 64.44\% & 78.91\% & 71.16\% & 87.62\% & 83.25\% & 94.34\% & 98.15\%\\
300 & 78.68\% & 64.24\% & 83.11\% & 71.39\% & 84.80\% & 78.08\% & 91.36\% & 87.02\% & 97.09\% & 98.61\%\\
\bottomrule
\end{tabular}}
\caption{Simulation results examining the proportion of rejections of the null hypothesis that the PTE is constant over time i.e., temporal homogeneity, for Scenario 3 (Null hypothesis is false) at an $\alpha$-level of 0.05}
\label{tab:simul7_case3}
\end{table}

\begin{table}
\centering
\resizebox{\textwidth}{!}{
\begin{tabular}{r @{\hskip 0.5cm} ll @{\hskip 0.5cm} ll @{\hskip 0.5cm} ll @{\hskip 0.5cm} ll @{\hskip 0.5cm} ll}
\toprule
Sample &  \multicolumn{2}{c}{1 Year} &  \multicolumn{2}{c}{2 Year} &  \multicolumn{2}{c}{3 Year} &  \multicolumn{2}{c}{5 Year} &  \multicolumn{2}{c}{10 Year}\\
size & \textbf{MSD} & \textbf{Wald} & \textbf{MSD} & \textbf{Wald} & \textbf{MSD} & \textbf{Wald} & \textbf{MSD} &\textbf{Wald} & \textbf{MSD} & \textbf{Wald}\\
\midrule
50 & 48.71\% & 41.09\% & 49.04\% & 51.06\% & 47.28\% & 65.89\% & 42.75\% & 94.07\% & 41.19\% & 100.00\%\\
100 & 73.74\% & 63.54\% & 71.82\% & 63.44\% & 67.78\% & 66.13\% & 63.51\% & 78.64\% & 58.05\% & 99.93\%\\
150 & 87.28\% & 79.93\% & 85.46\% & 75.26\% & 83.34\% & 75.43\% & 78.48\% & 78.34\% & 74.24\% & 97.65\%\\
200 & 94.14\% & 88.25\% & 93.28\% & 86.92\% & 91.85\% & 84.01\% & 88.01\% & 83.08\% & 84.80\% & 93.34\%\\
250 & 97.72\% & 95.33\% & 97.05\% & 91.82\% & 96.52\% & 89.97\% & 94.50\% & 87.55\% & 90.96\% & 92.65\%\\
\addlinespace
300 & 99.04\% & 97.12\% & 98.58\% & 95.20\% & 98.18\% & 94.04\% & 96.69\% & 90.33\% & 95.56\% & 93.61\%\\
\bottomrule
\end{tabular}}
\caption{Simulation results examining the proportion of rejections of the null hypothesis that the PTE is constant over time i.e., temporal homogeneity, for Scenario 4 (Null hypothesis is false) at an $\alpha$-level of 0.05}
\label{tab:simul7_case4}
\end{table}

\begin{table}
\centering
\resizebox{\textwidth}{!}{
\begin{tabular}{r @{\hskip 0.5cm} ll @{\hskip 0.5cm} ll @{\hskip 0.5cm} ll @{\hskip 0.5cm} ll @{\hskip 0.5cm} ll}
\toprule
Sample &  \multicolumn{2}{c}{1 Year} &  \multicolumn{2}{c}{2 Year} &  \multicolumn{2}{c}{3 Year} &  \multicolumn{2}{c}{5 Year} &  \multicolumn{2}{c}{10 Year}\\
size & \textbf{MSD} & \textbf{Wald} & \textbf{MSD} & \textbf{Wald} & \textbf{MSD} & \textbf{Wald} & \textbf{MSD} &\textbf{Wald} & \textbf{MSD} & \textbf{Wald}\\
\midrule
50 & 49.24\% & 41.26\% & 49.37\% & 49.40\% & 46.85\% & 64.44\% & 43.64\% & 92.98\% & 41.23\% & 100.00\%\\
100 & 76.89\% & 65.60\% & 74.83\% & 65.53\% & 72.58\% & 67.19\% & 67.12\% & 78.87\% & 60.60\% & 99.74\%\\
150 & 89.60\% & 82.81\% & 87.42\% & 78.15\% & 85.79\% & 76.99\% & 80.40\% & 79.44\% & 77.58\% & 97.22\%\\
200 & 96.19\% & 92.65\% & 95.70\% & 88.05\% & 93.25\% & 85.33\% & 90.79\% & 85.56\% & 87.58\% & 93.87\%\\
250 & 98.94\% & 96.75\% & 98.01\% & 94.24\% & 97.32\% & 91.39\% & 96.09\% & 89.74\% & 94.01\% & 93.41\%\\
\addlinespace
300 & 99.47\% & 98.68\% & 99.14\% & 97.22\% & 98.61\% & 95.43\% & 98.44\% & 93.38\% & 97.45\% & 94.30\%\\
\bottomrule
\end{tabular}}
\caption{Simulation results examining the proportion of rejections of the null hypothesis that the PTE is constant over time i.e., temporal homogeneity, for Scenario 5 (Null hypothesis is false) at an $\alpha$-level of 0.05}
\label{tab:simul7_case5}
\end{table}

\clearpage

\begin{figure}
\centering
    \begin{subfigure}[c]{0.59\textwidth}
\centering
    \includegraphics[scale=1]{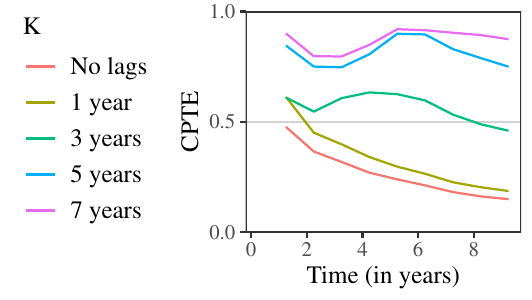}
    \caption{Estimation results using various lags}\label{fig:applied1a}
    \end{subfigure}
    \begin{subfigure}[c]{0.39\textwidth}
\centering
    \includegraphics[scale=1]{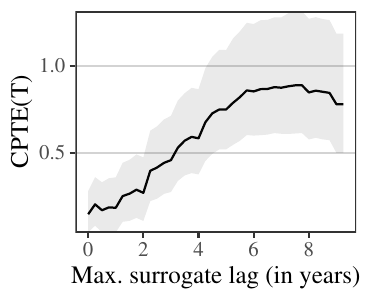}
    \caption{Estimated CPTE at time $T$}\label{fig:applied1b}
    \end{subfigure}
    \caption{Diabetes clinical trial results examining change in  hemoglobin A1c as a surrogate for change in albumin excretion rate; (a) each colored line corresponds to the estimated cumulative proportion of treatment effect explained (CPTE) using our proposed methods over time using the specified number of lags; a gray line is shown at a CPTE of 0.5 for reference; (a) versus using 0 lags; (b) the solid black line reflects the estimated CPTE at time $T$ (the final time) as a function of the maximum surrogate lags used in the corresponding model; the shaded region corresponds to estimated 90\% pointwise confidence intervals; a gray line is shown at a CPTE of 0.5 and 1.0 for reference.}
    \label{fig:applied1}
\end{figure}

\begin{table}
\centering
\begin{tabular}[t]{llrr}
\hline
\multirow{2}{5em}{Max. lag} & \multirow{2}{5em}{Point estimate} & \multicolumn{2}{c}  {Confidence interval}\\
  &  & Lower bound & Upper bound\\
\hline
No lags & 0.1498 & 0.0471 & 0.2831\\
1 year & 0.1868 & 0.0467 & 0.3614\\
3 years & 0.4595 & 0.2774 & 0.7130\\
5 years & 0.7497 & 0.5212 & 1.0900\\
7 years & 0.8741 & 0.6106 & 1.2763\\
9 years & 0.7795 & 0.5011 & 1.1838\\
\hline
\end{tabular}
\caption{Point estimates and 90\% confidence intervals for the cumulative proportion of treatment effect explained (CPTE) at the last time $T$ using different lag choices in the diabetes clinical trial examining change in  hemoglobin A1c as a surrogate for change in albumin excretion rate.}
\label{tab:applied_cpte_end}
\end{table}

\end{document}